\documentclass[structabstract]{aa}
\usepackage{graphicx}
\usepackage{txfonts}
\usepackage{longtable}
\usepackage{lscape}
\usepackage{natbib}
\begin{document}
   \title{New ultracool subdwarfs identified in large-scale surveys using Virtual 
Observatory tools \thanks{Based on observations 
made with ESO Telescopes at the La Silla Paranal Observatory under programme 
ID 084.C-0928A} \thanks{Based on observations made with the 
Nordic Optical Telescope, operated on the island of La Palma jointly by Denmark, 
Finland, Iceland, Norway, and Sweden, in the Spanish Observatorio del Roque de 
los Muchachos of the Instituto de Astrof\'isica de Canarias.}}

   \subtitle{Part I: UKIDSS LAS DR5 vs SDSS DR7}

   \author{N. Lodieu \inst{1,2}
          \and
          M. Espinoza Contreras \inst{1}
          \and
          M. R. Zapatero Osorio \inst{3}
          \and 
          E. Solano \inst{4,5}
          \and
          M. Aberasturi \inst{4,5}
          \and
          E. L. Mart\'in \inst{3}
          }

   \offprints{N. Lodieu}

   \institute{Instituto de Astrof\'isica de Canarias (IAC), Calle V\'ia L\'actea s/n, E-38200 La Laguna, Tenerife, Spain\\
         \email{nlodieu@iac.es, marcela@iac.es}
         \and
         Departamento de Astrof\'isica, Universidad de La Laguna (ULL),
E-38205 La Laguna, Tenerife, Spain
         \and
         Centro de Astrobiolog\'ia (CSIC-INTA), Ctra. Ajalvir km 4, 28850, Torrej\'on de Ardoz, Madrid, Spain \\
         \email{mosorio,ege@cab.inta-csic.es}
         \and
         Centro de Astrobiolog\'ia (INTA-CSIC), Departamento de Astrof\'isica. P.O. Box 78, E-28691 Villanueva de la Ca\~nada, Madrid, Spain \\
         \email{esm,miriam@cab.inta-csic.es}
         \and
         Spanish Virtual Observatory
             }

   \date{\today{}; \today{}}

% \abstract{}{}{}{}{} 
% 5 {} token are mandatory
 
  \abstract
   {}
   {The aim of the project is to improve our knowledge on the low-mass and 
   low-metallicity population to investigate the influence of metallicity on
   the stellar (and substellar) mass function.}
   {We present the results of a photometric and proper motion search aimed 
at unearthing ultracool subdwarfs in large-scale surveys. We employed and 
combined the Fifth Data Release (DR5) of the UKIRT Infrared Deep Sky Survey 
(UKIDSS) Large Area Survey (LAS) and the Sloan Digital Sky Survey (SDSS) Data 
Release 7 complemented with ancillary data from the Two Micron All-Sky Survey 
(2MASS), the DEep Near-Infrared Survey (DENIS) and the SuperCOSMOS Sky 
Surveys (SSS).}
   {The SDSS DR7 vs UKIDSS LAS DR5 search returned a total of 32 ultracool 
subdwarf candidates, only two being recognised as a subdwarf in the literature. 
Twenty-seven candidates, including the two known ones, were followed-up 
spectroscopically in the optical between 600 and 1000 nm thus covering strong 
spectral features indicative of low metallicity (e.g., CaH) , 21 
with the Very Large Telescope, one with the Nordic Optical Telescope, and five
were extracted from the Sloan spectroscopic database to assess (or refute) 
their low-metal content. We confirmed 20 candidates as subdwarfs, extreme 
subdwarfs or ultra-subdwarfs with spectral types later than M5; this represents 
a success rate of $\ge$60\%. Among those 20 new subdwarfs, we identified 
two early-L subdwarfs very likely located within 100 pc that we propose as 
templates for future searches because they are the first examples of their 
subclass. Another seven sources are solar-metallicity M dwarfs with spectral 
types between M4 and M7 without H$\alpha$ emission, suggesting that they are 
old M dwarfs. The remaining five candidates do not have spectroscopic 
follow-up yet; only one remains as a bona-fide ultracool subdwarf
after revision of their proper motions.
We assigned spectral types based on the current classification 
schemes and, when possible, we measured their radial velocities.
Using the limited number of subdwarfs with trigonometric parallaxes, 
we estimated distances ranging from $\sim$95 to $\sim$600 pc for the new 
subdwarfs. We provide mid-infrared photometry extracted from the WISE 
satellite databases for two subdwarfs and discuss their colours.
Finally, we estimate a lower limit of the surface density of ultracool 
subdwarfs of the order of 5000--5700 times lower than that of
solar-metallicity late-M dwarfs.}
   {}


   \titlerunning{Ultracool subdwarfs in large-scale surveys using VO tools}
   \maketitle
%
%________________________________________________________________

%
%%%%%%%%%%%%%%%%%%%%%%%%%%
%%%%% Introduction %%%%%
%%%%%%%%%%%%%%%%%%%%%%%%%%
%
\section{Introduction}
\label{subdw:intro}

Cool subdwarfs are metal-deficient population II dwarfs which appear 
less luminous than their solar-metallicity counterparts due to the dearth 
of metals in their atmospheres \citep{baraffe97}. They tend to exhibit halo 
or thick disk kinematics, including noticeable proper motion and large 
heliocentric velocities \citep{gizis97a}. They are very old and represent 
useful tracers of the Galactic chemical history \citep{burgasser03b}. 
The adopted classification for M subdwarfs (sdM) and extreme subdwarfs (esdM) 
has recently been revised by \citet{lepine07c}. A new class of subdwarfs, the 
ultra-subdwarfs (usdM), has been added to the sdM and esdM classes originally 
defined by \citet{gizis97a}. The new scheme is based on a parameter, 
$\zeta_{\rm TiO/CaH}$, which quantifies 
the weakening of the strength of the TiO band (in the optical) as a function 
of metallicity. An alternative classification based on temperature, gravity,
and metallicity has been proposed by \citet{jao08}. The range of metallicity
for subdwarfs, extreme subdwarfs, and ultra-subdwarfs span
approximately $-$0.5 and $-$1.0, $-$1.0 and $-$1.5 and below $-$1.5, 
respectively \citep{gizis97a,woolf09}. M-type subdwarfs have typically 
effective temperatures below $\sim$3500--4000\,K 
\citep[depending on the metallicity][]{baraffe97,woolf09} and should display 
high gravity ($\log$\,g$\sim$5.5) although some variations is seen among 
low-metallicity spectra \citep{jao08}.

Subdwarfs were generally identified from optical ($B_{J}$, $R$, and $I$)
proper motion catalogues on photographic plates at different epochs
\citep{luyten79,luyten80,scholz00,lepine03a,lodieu05b}.
Several surveys have been conducted to search for subdwarfs
over a large temperature range, including hot \citep{ryan89},
intermediate \citep{yong03,digby03}, and
cool components \citep{gizis97a}. A growing number of M subdwarfs have
been announced over the past years, raising the number of metal-deficient
dwarfs with spectral types later than M7 to about ten
\citep{gizis97a,gizis97b,schweitzer99,lepine03b,scholz04b,cushing09}.
This number increased significantly after the discovery of 23 late-type
subdwarfs in the Sloan Digital Sky Survey \citep[SDSS;][]{adelman_mccarthy09} 
spectroscopic database \citep{lepine08b} and 15 others in the multi-epoch
database of the 2MASS survey \citep{kirkpatrick10}. Moreover, the 
hydrogen-burning limit has been crossed with the discovery of the first 
substellar subdwarf in the 2MASS database \citep{burgasser03b}. This discovery 
was quickly followed by the announcement of another L subdwarf 
\citep{burgasser04} and, more recently, by seven new ones
\citep{sivarani09,cushing09,lodieu10a,kirkpatrick10,schmidt10a,bowler10a}.
This number of ultracool subdwarfs remains however
very small and is at odds with the numerous L and T dwarfs
reported in the solar neighbourhood within the framework
of large-scale optical and infrared surveys, including the
Two Micron All Sky Survey \citep[2MASS; e.g.][]{kirkpatrick00,burgasser02},
DEep Near Infrared Survey \citep[DENIS; e.g.][]{delfosse97,delfosse99,martin99a},
SDSS \citep[e.g.][]{fan00,leggett00,geballe02}, the UKIRT Infrared Deep Sky 
Survey \citep[UKIDSS; e.g.][]{lodieu07b,pinfield08,burningham10b}, 
the Canada-France-Hawaii Brown Dwarf survey 
\citep{delorme08a,reyle10,albert11}, and WISE \citep{kirkpatrick11,cushing11}.

The aim of our study is to identify a complete census of metal-poor dwarfs
to bridge the gap between the coolest M subdwarfs and the recent L subdwarfs
identified in various surveys.
In this paper we report the outcome of a photometric and proper motion
searches using the UKIRT Infrared Deep Sky Survey \citep[UKIDSS;][]{lawrence07} 
Large Area Survey (LAS) Data Release 5 (DR5) and the Sloan Digital Sky Survey 
(SDSS) Data Release 7 \citep[DR7;][]{abazajian09}. This catalogue search was 
complemented by ancillary data from 2MASS \citep{cutri03,skrutskie06}, DENIS
\citep{DENISconsortium2005}, and SuperCOSMOS \citep{hambly01a,hambly01b,hambly01c}.
In Sect.\ \ref{subdw:select} we describe the photometric and proper motion 
criteria designed to identify ultracool subdwarfs in public databases using
Virtual Observatory tools. 
In Sect.\ \ref{subdw:spec_obs} we detail the spectroscopic follow-up conducted 
at optical wavelengths with the visual and near UV FOcal Reducer and low 
dispersion Spectrograph \citep[FORS2;][]{appenzeller98} mounted on the Very 
Large Telescope (VLT), the ALFOSC spectrograph on the Nordic Optical Telescope 
(NOT) and complemented by optical spectra downloaded from the SDSS 
spectroscopic database.
In Sect.\ \ref{subdw:subdw} we present the results of the spectroscopic 
analysis and determine the main properties of the newly identified ultracool 
subdwarfs including colours, spectral types. radial velocities, and distances. 
Finally, we summarise our work in Sect.\ \ref{subdw:summary}.

%
%%%%%%%%%%%%%%%%%%%%%%%%%%%%%%%%%%%%%%%%%%
%%%%%%%% (J-K,i-J) CCD %%%%%%%%%%%
%%%%%%%%%%%%%%%%%%%%%%%%%%%%%%%%%%%%%%%%%%
%
\begin{figure}
  \centering
  \includegraphics[width=\linewidth, angle=0]{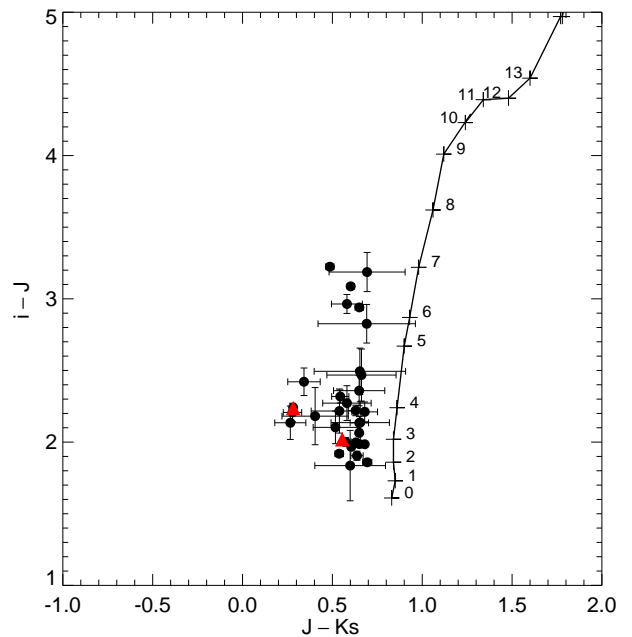}
  \caption{($J-K_{s}$,$i-J$) colour-colour diagram showing the positions of 
our subdwarf candidates identified in the SDSS/UKIDSS search (filled circles)
with their respective error bars. Overplotted as filled red triangles are 
two known ultracool subdwarfs from \citet{lepine08b} with $J$ and $K$ 
photometry from UKIDSS LAS DR5 (error bars are smaller than the symbols) 
and the solar-metallicity M/L dwarf sequence (line with crosses; spectral 
types are labelled; 0$\equiv$M0, 10$\equiv$L0) from \citet{west08} and 
\citet{schmidt10b}. The photometry shown on this diagram in the MKO 
photometric system \citep{hewett06}.
}
  \label{fig_subdw:fig_ccd1}
\end{figure}
%

%
%%%%%%%%%%%%%%%%%%%%%%%%%%%%%%%%%%%%%%%%%%
%%%%%%%%%% Sample selection %%%%%%%%%
%%%%%%%%%%%%%%%%%%%%%%%%%%%%%%%%%%%%%%%%%%
%
\section{Sample selection}
\label{subdw:select}

In this section we describe the selection procedure developed to unveil 
new subdwarfs in the cross-correlation of the SDSS DR7 and UKIDSS LAS DR5
catalogues.

The idea of a generic photometric search for subdwarfs was triggered by the 
($I-J$,$J-K$) colour-colour diagram is presented in Fig.\ 4 of \citet{scholz04b}. 
A similar diagram is shown in Fig.\ \ref{fig_subdw:fig_ccd1}. Known ultracool 
subdwarfs identified by \citet{lepine08b} in the SDSS spectroscopic database are 
plotted along with the sequence of field M/L dwarfs \citep{west08,schmidt10b}.
Subdwarfs tend to be bluer than solar metallicity M/L dwarfs but occupy the 
same region in the $i-J$ domain as their solar metallicity counterparts, making 
their identification difficult in a pure optical-to-infrared search. However, 
the onset of collision-induced molecular hydrogen H$_{2}$ opacity at 
near-infrared wavelengths generates bluer $J-H$ and $J-K_{s}$ colours 
($J-K_{s} \leq$ 0.7) than observed for solar metallicity M and L dwarfs. 

\begin{figure*}
  \centering
  \includegraphics[width=0.48\linewidth, angle=0]{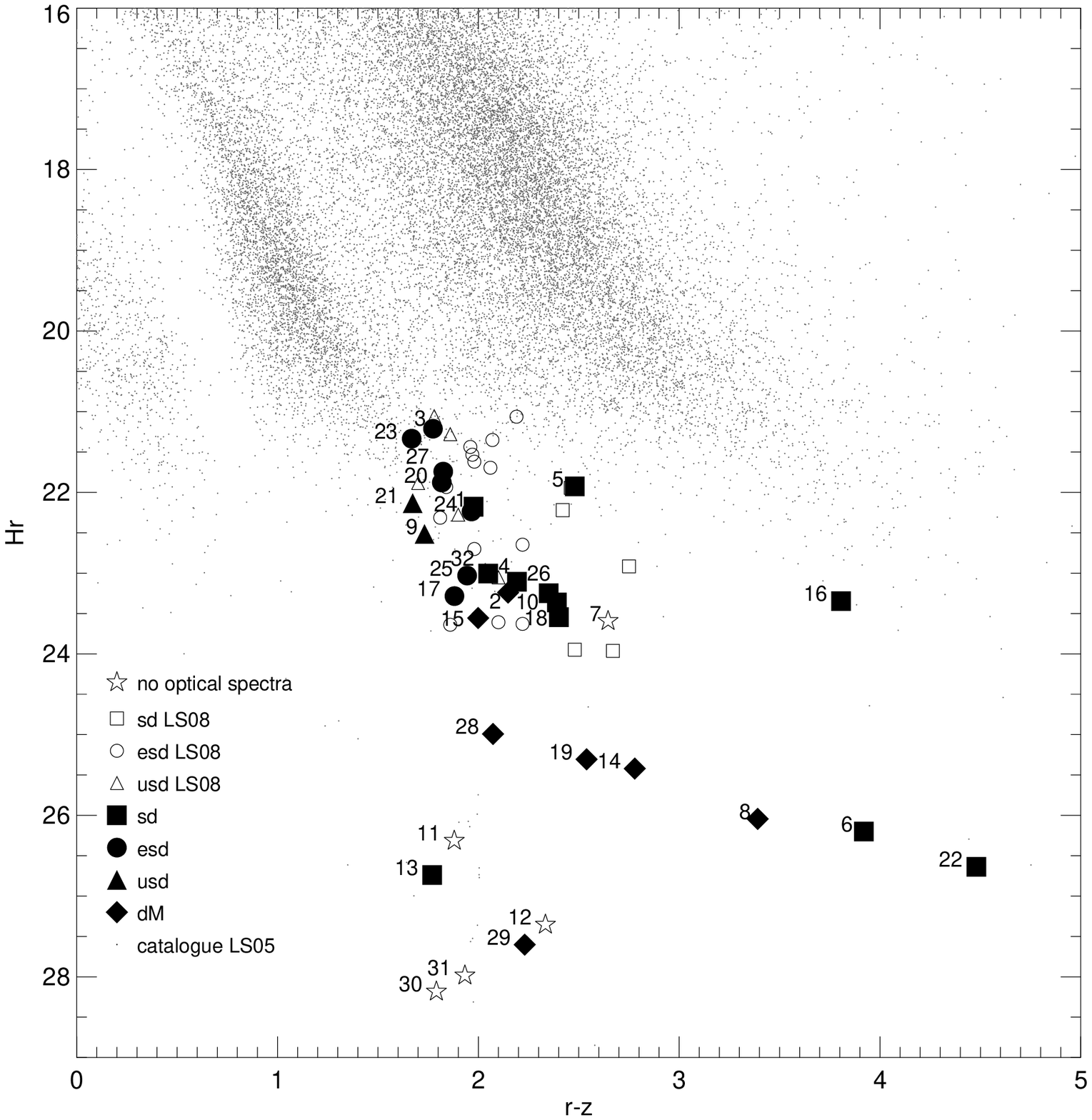}
  \includegraphics[width=0.48\linewidth, angle=0]{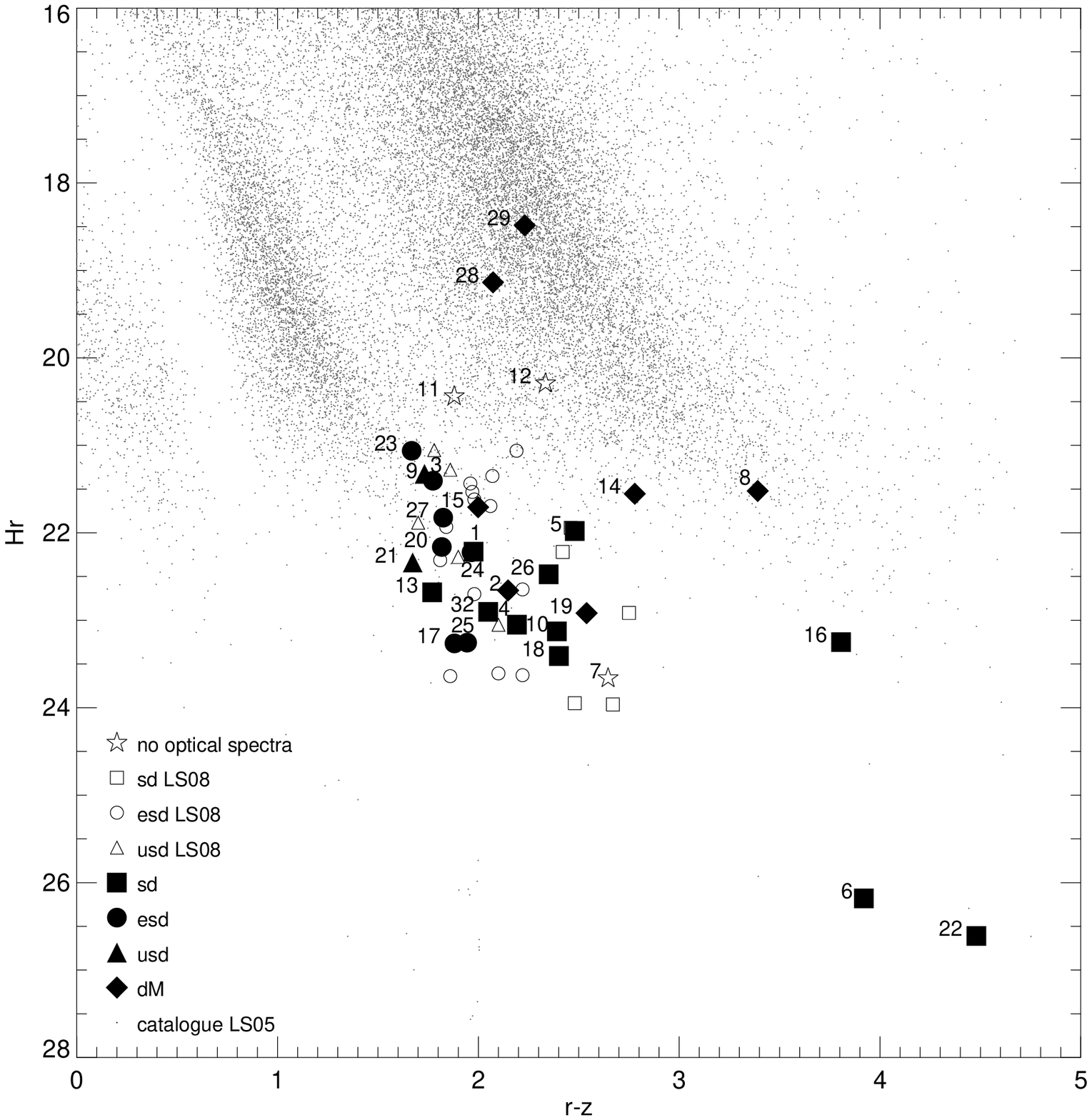}
  \caption{
Reduced proper motion diagram for new ultracool subdwarfs identified in 
the SDSS DR7 vs UKIDSS LAS DR5 search. The small dots represent all sources 
in the catalogue published by \citet[LS05;][]{lepine05d} with counterpart
in the SDSS DR7 database. Our new subdwarfs, extreme subdwarfs, and 
ultra-subdwarfs are marked as filled squares, circles, and triangles, 
respectively. Diamonds represent our candidates classified as 
solar-metallicity M dwarfs. The five sources without optical 
spectroscopy are marked with a star symbol. Known subdwarfs from the 
literature are marked as open symbols \citep[LS08;][]{lepine08b}.
{\it{Left:}} Diagram for the values given by the Virtual Observatory.
{\it{Right:}} Diagram using the revised proper motions computed with
the method detailed in Section \ref{subdw:revised_PM}.
}
  \label{fig_subdw:fig_RPM}
\end{figure*}

To optimise our photometric selection, we have employed the
reduced proper motion (H$_{r}$=$r$\,$+$\,5$\times$\,$\log$($\mu$)\,$+$\,5) 
as a proxy for metallicity. This parameter is useful to 
separate solar-metallicity stars from subdwarfs, and white dwarfs 
\citep{jones72a,evans92,salim02,lepine05d,burgasser07b,lodieu09a}. A reduced 
proper motion diagram is shown in Fig.\ \ref{fig_subdw:fig_RPM} with known 
subdwarfs from \citet{lepine08b} marked as open symbols and our new candidates 
confirmed as subdwarfs shown as filled symbols.

%
%%%%%%%%%%%%%%%%%%%%%%%%%%%%%%%%%%%%%%%%%%
%%%%% Landscape Table: SDSS/UKIDSS %%%%%
%%%%%%%%%%%%%%%%%%%%%%%%%%%%%%%%%%%%%%%%%%
%
\longtab{1}{\scriptsize
\begin{landscape}
  \begin{longtable}{| @{\hspace{1mm}}c @{\hspace{1mm}}c | @{\hspace{1mm}}c @{\hspace{2mm}}c@{\hspace{1mm}} | @{\hspace{1mm}}c @{\hspace{2mm}}c @{\hspace{2mm}}c @{\hspace{2mm}}c @{\hspace{2mm}}c @{\hspace{2mm}}c@{\hspace{1mm}} | @{\hspace{1mm}}c @{\hspace{2mm}}c @{\hspace{2mm}}c @{\hspace{2mm}}c @{\hspace{2mm}}c@{\hspace{1mm}} | @{\hspace{1mm}}c @{\hspace{2mm}}c@{\hspace{1mm}} |}
  \caption{\label{tab_subdw:tab_SDSS_UKIDSS} Running number, names according
to the IAU nomenclature of UKIDSS, coordinates (in J2000), SDSS optical 
($ugriz$) magnitudes with epoch, UKIDSS 
near-infrared ($YJHK$) magnitudes with epochs, total 
proper motion ($\mu$ in arcsec/yr), and the reduced proper motion (H$_{r}$) of
ultracool subdwarf candidates extracted from the cross-match between the UKIDSS
LAS DR5 and SDSS DR7} \\
  \hline
  \hline
ID & Name & R.A. & Dec & SDSS$u$ & SDSS$g$ & SDSS$r$ & SDSS$i$ & SDSS$z$ & Epoch & $Y$ & $J$ & $H$ & $K$ & Epoch & $\mu$ & H$_{r}$ \\
   &      & hh:mm:ss.ss & $^{\circ}$:$'$:$''$ & mag & mag & mag & mag & mag & yr & mag & mag & mag & mag & yr & ``/yr & mag \\
  \hline
  \endfirsthead
  \caption{continued.} \\
  \hline
  \hline
ID & Name & R.A. & Dec & SDSS$u$ & SDSS$g$ & SDSS$r$ & SDSS$i$ & SDSS$z$ & Epoch & $Y$ & $J$ & $H$ & $K$ & Epoch & $\mu$ & H$_{r}$ \\
   &      & hh:mm:ss.ss & $^{\circ}$:$'$:$''$ & mag & mag & mag & mag & mag & yr & mag & mag & mag & mag & yr & ``/yr & mag \\
  \hline
  \endhead
  \hline
  \endfoot
 1 & ULAS J004539.97+135032.7 & 00:45:39.97 & $+$13:50:32.7 & 24.559$\pm$1.069 & 25.657$\pm$0.771 & 21.500$\pm$0.075 & 20.254$\pm$0.040 & 19.524$\pm$0.079 & 1999.78 & 18.755$\pm$0.055 & 18.043$\pm$0.059 & 17.630$\pm$0.088 & 17.363$\pm$0.100 & 2007.71 &  0.14 & 22.177 \cr 
 2 & ULAS J012830.89+134507.4 & 01:28:30.89 & $+$13:45:07.4 & 24.930$\pm$1.485 & 26.014$\pm$0.640 & 22.595$\pm$0.294 & 21.154$\pm$0.121 & 20.447$\pm$0.256 & 1999.78 & 19.693$\pm$0.172 & 19.318$\pm$0.155 & 18.585$\pm$0.145 & 18.719$\pm$0.301 & 2007.64 &  0.13 & 23.240 \cr 
 3 & ULAS J013346.25+132822.4 & 01:33:46.25 & $+$13:28:22.4 & 23.464$\pm$0.835 & 20.822$\pm$0.035 & 18.955$\pm$0.012 & 17.822$\pm$0.008 & 17.181$\pm$0.013 & 1999.78 & 16.351$\pm$0.009 & 15.855$\pm$0.010 & 15.467$\pm$0.015 & 15.250$\pm$0.017 & 2007.63 &  0.28 & 21.211$^{a}$ \cr 
 4 & ULAS J015034.33+142002.4 & 01:50:34.33 & $+$14:20:02.4 & 23.456$\pm$0.617 & 22.912$\pm$0.196 & 21.087$\pm$0.058 & 19.645$\pm$0.026 & 18.895$\pm$0.040 & 1999.78 & 18.018$\pm$0.037 & 17.424$\pm$0.032 & 16.946$\pm$0.042 & 16.794$\pm$0.069 & 2007.64 &  0.25 & 23.106 \cr 
 5 & ULAS J02 533.75+123824.1 & 02:05:33.75 & $+$12:38:24.1 & 23.281$\pm$0.597 & 21.944$\pm$0.104 & 19.767$\pm$0.021 & 18.113$\pm$0.010 & 17.287$\pm$0.018 & 2001.89 & 16.456$\pm$0.008 & 15.872$\pm$0.009 & 15.709$\pm$0.012 & 15.590$\pm$0.018 & 2006.89 &  0.27 & 21.925 \cr 
 6 & ULAS J033350.84+001406.1 & 03:33:50.84 & $+$00:14:06.1 & 23.392$\pm$0.753 & 23.818$\pm$0.374 & 21.731$\pm$0.102 & 19.193$\pm$0.017 & 17.811$\pm$0.021 & 1999.78 & 16.814$\pm$0.006 & 16.106$\pm$0.010 & 15.765$\pm$0.009 & 15.504$\pm$0.019 & 2007.63 &  0.78 & 26.201 \cr 
 7 & ULAS J083525.41-000940.5 & 08:35:25.41 & $-$00:09:40.5 & 24.659$\pm$0.781 & 24.601$\pm$0.532 & 22.497$\pm$0.151 & 20.680$\pm$0.047 & 19.851$\pm$0.088 & 1999.22 & 19.098$\pm$0.057 & 18.259$\pm$0.077 & 18.186$\pm$0.099 & 17.917$\pm$0.138 & 2006.93 &  0.17 & 23.592 \cr 
 8 & ULAS J084153.89+020615.1 & 08:41:53.89 & $+$02:06:15.1 & 23.580$\pm$0.904 & 26.662$\pm$0.390 & 23.681$\pm$0.697 & 21.821$\pm$0.253 & 20.290$\pm$0.222 & 2000.92 & 19.638$\pm$0.094 & 18.995$\pm$0.097 & 18.496$\pm$0.114 & 18.304$\pm$0.194 & 2006.99 &  0.30 & 26.044 \cr 
 9 & ULAS J084358.50+060038.6 & 08:43:58.50 & $+$06:00:38.6 & 22.352$\pm$0.295 & 19.711$\pm$0.015 & 17.815$\pm$0.006 & 16.717$\pm$0.005 & 16.083$\pm$0.008 & 2002.19 & 15.268$\pm$0.003 & 14.732$\pm$0.004 & 14.307$\pm$0.004 & 14.052$\pm$0.006 & 2006.96 &  0.50 & 21.307$^{b}$ \cr 
10 & ULAS J085500.12+000204.1 & 08:55:00.12 & $+$00:02:04.1 & 24.020$\pm$0.725 & 24.205$\pm$0.400 & 22.320$\pm$0.128 & 20.745$\pm$0.048 & 19.929$\pm$0.107 & 2000.17 & 19.293$\pm$0.072 & 18.610$\pm$0.072 & 18.366$\pm$0.094 & 18.344$\pm$0.211 & 2006.93 &  0.16 & 23.364 \cr 
11 & ULAS J093502.10+8 551.8 & 09:35:02.10 & $+$08:05:51.8 & 23.088$\pm$0.548 & 24.742$\pm$0.709 & 22.374$\pm$0.185 & 21.318$\pm$0.121 & 20.494$\pm$0.197 & 2002.19 & 19.767$\pm$0.108 & 19.182$\pm$0.108 & 18.685$\pm$0.109 & 18.527$\pm$0.194 & 2007.06 &  0.61 & 26.314 \cr 
12 & ULAS J095133.35+070208.9 & 09:51:33.35 & $+$07:02:08.9 & 25.254$\pm$0.829 & 24.666$\pm$0.677 & 22.335$\pm$0.178 & 20.994$\pm$0.092 & 20.001$\pm$0.131 & 2002.93 & 19.384$\pm$0.063 & 18.891$\pm$0.083 & 18.311$\pm$0.088 & 18.374$\pm$0.175 & 2008.13 &  1.01 & 27.353 \cr 
13 & ULAS J100126.29-013426.6 & 10:01:26.29 & $-$01:34:26.6 & 25.659$\pm$0.956 & 24.326$\pm$0.700 & 22.214$\pm$0.170 & 21.051$\pm$0.107 & 20.444$\pm$0.206 & 2000.17 & 19.517$\pm$0.088 & 18.779$\pm$0.082 & 18.469$\pm$0.117 & 18.198$\pm$0.199 & 2005.99 &  0.80 & 26.739 \cr 
14 & ULAS J100743.96-022830.0 & 10:07:43.96 & $-$02:28:30.0 & 25.722$\pm$0.798 & 25.124$\pm$0.871 & 23.240$\pm$0.440 & 21.528$\pm$0.142 & 20.461$\pm$0.196 & 2000.17 & 19.891$\pm$0.098 & 19.060$\pm$0.130 & 18.534$\pm$0.106 & 18.398$\pm$0.227 & 2005.97 &  0.27 & 25.421 \cr 
15 & ULAS J101613.89+11311.4 & 10:16:13.89 & $+$01:13:11.4 & 24.805$\pm$1.451 & 24.590$\pm$0.988 & 22.513$\pm$0.260 & 21.036$\pm$0.113 & 20.514$\pm$0.261 & 2000.98 & 19.573$\pm$0.111 & 18.855$\pm$0.146 & 18.447$\pm$0.115 & 18.451$\pm$0.231 & 2008.13 &  0.16 & 23.556 \cr 
16 & ULAS J115826.62+044746.8 & 11:58:26.62 & $+$04:47:46.8 & 24.243$\pm$0.864 & 23.832$\pm$0.338 & 21.961$\pm$0.093 & 19.426$\pm$0.017 & 18.155$\pm$0.022 & 2001.14 & 17.121$\pm$0.014 & 16.486$\pm$0.016 & 16.124$\pm$0.014 & 15.836$\pm$0.020 & 2008.13 &  0.19 & 23.346 \cr 
17 & ULAS J120214.62+073113.8 & 12:02:14.62 & $+$07:31:13.8 & 24.028$\pm$0.953 & 22.681$\pm$0.130 & 20.601$\pm$0.034 & 19.444$\pm$0.018 & 18.720$\pm$0.038 & 2003.25 & 18.059$\pm$0.046 & 17.539$\pm$0.029 & 17.041$\pm$0.033 & 16.902$\pm$0.051 & 2007.29 &  0.34 & 23.284$^{c}$ \cr 
18 & ULAS J1215 8.37+040200.5 & 12:15:08.37 & $+$04:02:00.5 & 24.584$\pm$0.971 & 23.636$\pm$0.381 & 21.517$\pm$0.063 & 20.016$\pm$0.028 & 19.116$\pm$0.047 & 2001.14 & 18.302$\pm$0.039 & 17.698$\pm$0.039 & 17.242$\pm$0.046 & 17.154$\pm$0.076 & 2007.26 &  0.25 & 23.545 \cr 
19 & ULAS J122145.28+080404.4 & 12:21:45.28 & $+$08:04:04.4 & 26.002$\pm$0.968 & 26.215$\pm$0.756 & 22.897$\pm$0.430 & 21.635$\pm$0.227 & 20.358$\pm$0.235 & 2003.25 & 19.973$\pm$0.087 & 19.141$\pm$0.115 & 18.417$\pm$0.094 & 18.488$\pm$0.217 & 2007.06 &  0.30 & 25.306 \cr 
20 & ULAS J123659.43-002158.2 & 12:36:59.43 & $-$00:21:58.2 & 24.089$\pm$0.876 & 21.392$\pm$0.050 & 19.563$\pm$0.015 & 18.404$\pm$0.010 & 17.745$\pm$0.018 & 1999.22 & 16.900$\pm$0.016 & 16.406$\pm$0.012 & 16.040$\pm$0.015 & 15.775$\pm$0.023 & 2005.44 &  0.29 & 21.877$^{d}$ \cr 
21 & ULAS J124234.62+143306.2 & 12:42:34.62 & $+$14:33:06.2 & 25.128$\pm$0.972 & 21.662$\pm$0.059 & 19.794$\pm$0.020 & 18.740$\pm$0.014 & 18.121$\pm$0.028 & 2003.08 & 17.334$\pm$0.017 & 16.821$\pm$0.019 & 16.485$\pm$0.023 & 16.283$\pm$0.036 & 2007.26 &  0.29 & 22.133 \cr 
22 & ULAS J124425.90+102441.9 & 12:44:25.90 & $+$10:24:41.9 & 25.453$\pm$0.726 & 25.472$\pm$0.566 & 22.499$\pm$0.157 & 19.483$\pm$0.020 & 18.019$\pm$0.019 & 2002.19 & 16.982$\pm$0.008 & 16.259$\pm$0.010 & 15.999$\pm$0.011 & 15.773$\pm$0.019 & 2007.12 &  0.67 & 26.638 \cr 
23 & ULAS J124621.90+044309.9 & 12:46:21.90 & $+$04:43:09.9 & 23.343$\pm$0.639 & 21.912$\pm$0.095 & 20.083$\pm$0.024 & 19.000$\pm$0.016 & 18.415$\pm$0.041 & 2001.29 & 17.664$\pm$0.018 & 17.141$\pm$0.020 & 16.718$\pm$0.024 & 16.447$\pm$0.035 & 2007.05 &  0.18 & 21.334 \cr 
24 & ULAS J125635.91-001944.9 & 12:56:35.91 & $-$00:19:44.9 & 24.973$\pm$0.772 & 21.414$\pm$0.045 & 19.484$\pm$0.014 & 18.231$\pm$0.009 & 17.518$\pm$0.015 & 1999.22 & 16.694$\pm$0.009 & 16.167$\pm$0.009 & 15.729$\pm$0.011 & 15.518$\pm$0.017 & 2005.45 &  0.36 & 22.236$^{e}$ \cr 
25 & ULAS J131705.66+091016.9 & 13:17:05.66 & $+$09:10:16.9 & 25.781$\pm$0.946 & 22.569$\pm$0.147 & 20.505$\pm$0.038 & 19.313$\pm$0.017 & 18.561$\pm$0.032 & 2003.32 & 17.843$\pm$0.025 & 17.311$\pm$0.027 & 16.940$\pm$0.034 & 16.738$\pm$0.049 & 2006.47 &  0.32 & 23.030 \cr 
26 & ULAS J141806.71+000035.5 & 14:18:06.71 & $+$00:00:35.5 & 25.359$\pm$0.824 & 24.845$\pm$1.117 & 21.872$\pm$0.102 & 20.286$\pm$0.041 & 19.522$\pm$0.085 & 1999.22 & 18.595$\pm$0.055 & 18.079$\pm$0.031 & 17.723$\pm$0.090 & 17.801$\pm$0.183 & 2005.45 &  0.19 & 23.248 \cr 
27 & ULAS J145441.42+123556.7 & 14:54:41.42 & $+$12:35:56.7 & 22.972$\pm$0.350 & 21.191$\pm$0.035 & 19.274$\pm$0.012 & 18.107$\pm$0.009 & 17.449$\pm$0.013 & 2003.47 & 16.664$\pm$0.007 & 16.121$\pm$0.010 & 15.695$\pm$0.012 & 15.471$\pm$0.013 & 2007.26 &  0.31 & 21.742$^{f}$ \cr 
28 & ULAS J151211.64+064251.3 & 15:12:11.64 & $+$06:42:51.3 & 25.525$\pm$1.651 & 24.856$\pm$1.050 & 22.424$\pm$0.284 & 21.189$\pm$0.118 & 20.351$\pm$0.208 & 2003.32 & 19.501$\pm$0.078 & 18.830$\pm$0.080 & 18.503$\pm$0.109 & 18.181$\pm$0.160 & 2006.43 &  0.33 & 24.992 \cr 
29 & ULAS J154331.93+24537.8 & 15:43:31.93 & $+$02:45:37.8 & 25.682$\pm$0.934 & 25.591$\pm$0.715 & 23.278$\pm$0.383 & 21.929$\pm$0.187 & 21.047$\pm$0.366 & 2000.36 & 19.536$\pm$0.092 & 18.742$\pm$0.099 & 18.160$\pm$0.124 & 18.049$\pm$0.194 & 2006.57 &  0.73 & 27.602 \cr 
30 & ULAS J154450.60+051613.6 & 15:44:50.60 & $+$05:16:13.6 & 23.830$\pm$1.103 & 24.671$\pm$0.770 & 22.769$\pm$0.192 & 21.449$\pm$0.114 & 20.978$\pm$0.407 & 2001.46 & 19.685$\pm$0.117 & 19.232$\pm$0.108 & 18.604$\pm$0.127 & 18.693$\pm$0.214 & 2005.43 &  1.21 & 28.182 \cr 
31 & ULAS J154855.74+080508.2 & 15:48:55.74 & $+$08:05:08.2 & 24.355$\pm$1.431 & 25.155$\pm$0.720 & 22.719$\pm$0.214 & 21.178$\pm$0.070 & 20.786$\pm$0.241 & 2003.25 & 18.783$\pm$0.045 & 18.214$\pm$0.050 & 17.745$\pm$0.083 & 17.633$\pm$0.127 & 2006.44 &  1.13 & 27.983 \cr 
32 & ULAS J233359.39+04935.2 & 23:33:59.39 & $+$00:49:35.2 & 24.651$\pm$0.928 & 23.284$\pm$0.200 & 21.476$\pm$0.066 & 20.152$\pm$0.036 & 19.427$\pm$0.063 & 2001.79 & 18.541$\pm$0.058 & 18.018$\pm$0.036 & 17.628$\pm$0.070 & 17.368$\pm$0.090 & 2007.61 &  0.20 & 23.005$^{g}$ \cr 
\hline
\end{longtable}
% \footnote{Notes on individual objects}
Notes: $^{a}$ LP\,468-277 \citep{lepine05d,west08} \\
$^{b}$ LHS\,2045 (ID=9) \citep{lepine05d,west08} \\
$^{c}$ SDSS J120214.61$+$073113.7 (ID=17) classified as M4 by \citet{west08} \\
$^{d}$ 2MASS J12365942$-$0021578 (ID=2) classified as M3 by \citet{west08} \\
$^{e}$ 2MASS J12563590$-$00194507 (ID=24) classified as M3 by \citet{west08} \\
$^{f}$ 2MASS J14544145$+$1235576 \citep[ID=27;][]{lepine05d}  \\
$^{g}$ Located at $\sim$42 arcsec from 2MASS J23335840+0050119 classified
as a L0 dwarf by \citet{zhang10} 
\end{landscape}
% End \longtab
%
}

%
%%%%%%%%%%%%%%%%%%%%%%%%%%%%%%%
%%%%% RPM diagram %%%%%
%%%%%%%%%%%%%%%%%%%%%%%%%%%%%%%
%

Our search has been conducted taking 
advantage of Virtual Observatory (VO)\footnote{The Virtual Observatory
(http://www.ivoa.net) is an operational research infrastructure designed
to facilitate the access and analysis of the information hosted in 
astronomical archives} tools, 
namely STILTS\footnote{http://www.star.bris.ac.uk/$\sim$mbt/stilts/} and 
Aladin\footnote{http://aladin.u-strasbg.fr/} \citep{bonnarel00}.
The detailed photometric and proper motion criteria used for the SDSS DR7 vs
UKIDSS LAS DR5 search are given below. The resulting total number of 
candidates returned by this query is 33 
(Table \ref{tab_subdw:tab_SDSS_UKIDSS}) but one was rejected after looking 
at the images. Those candidates are shown as 
filled symbols in Figs.\ \ref{fig_subdw:fig_ccd1} and \ref{fig_subdw:fig_RPM}.
The finding charts created with Aladin are displayed in 
Figs.\ \ref{fig_subdw:fc_subdw_1} and \ref{fig_subdw:fc_subdw_2} of the 
Appendix.

\begin{itemize}
\item Only SDSS point sources (class=6) were considered
\item UKIDSS point sources ({\tt{mergedClass}} parameter equal to $-$1 or 
$-$2) and detections in $Y$, $J$, and $H$
\item Only sources fainter than $Y$ $>$ 10.5, $J$ $>$ 10.5, and $H$ $>$ 10.2 mag
were considered to avoid saturated stars
\item For all SDSS sources, we looked for UKIDSS counterparts between 1 and 
5 arcsec, implying that we are sensitive to dwarfs with proper motions between 
0.125 and $\sim$1.0 arcsec per year depending on the baseline existing between 
SDSS and UKIDSS (our targets span the 3.11--7.93 year baseline)
\item Colour selection were then applied as follows: $r-i \geq$ 1.0 mag, 
$g-r \geq$ 1.8 mag, $r-z \geq$ 1.6 mag
\item Only UKIDSS sources with good quality flags were kept i.e.\ {\tt{ppErrBits}} parameter in each filter less or equal to 256
\item Infrared criterion of $J-K \leq$ 0.7 mag was applied
\item {\tt{Xi}} and {\tt{Eta}} parameters refering to positional matching should 
be between $-$0.5 and 0.5
\item Only objects satisfying H$_{r} \geq$ 20.7 mag were kept
\end{itemize}

The proper motion of each candidate was computed using the UKIDSS LAS and SDSS
positions only but images from 2MASS, DENIS, and SuperCOSMOS were checked by 
eye for additional epochs to get rid off false positives. A revision
of the proper motions is detailed in Section \ref{subdw:revised_PM_results}.

Candidates were visually inspected using the scripting capabilities of Aladin. 
Sources from DENIS, 2MASS, SDSS, SuperCOSMOS and UKIDSS as well images from 
UKIDSS, 2MASS and SDSS were used in the analysis. False candidates were 
rejected due to several reasons, the most likely being the mismatch between 
the SDSS source and the associated UKIDSS counterpart.

We imposed a lower limit on the proper motion on purpose in order to bias our
search towards halo objects \citep[e.g.,][]{scholz00} and avoid contamination 
by ``normal'' dwarfs and extragalactic sources. Originally, we took 0.5 arcsec
(instead of 1 arcsec) as the lower limit for the SDSS-UKIDSS separation but it 
produced a large number of false candidates. Moreover, imposing a detection
in the $g$-band is hampering the detection of cooler subdwarfs which would be
faint at short optical wavelengths. We tested our completeness limit by 
checking how many of the $\sim$250 known ultracool subdwarfs from different 
studies are recovered by our work
\citep{gizis97a,schweitzer99,lepine02,lepine03a,scholz04b,scholz04c,lodieu05b,reid05,lepine08b,kirkpatrick10}.
Of the $\sim$250 M- and L-type subdwarfs, only 11 lie in the common area 
between UKIDSS LAS DR5 and SDSS DR7:
seven do not satisfy either the optical colour constraints because they have
spectral types earlier than M5 or are missing near-infrared photometry in at
least one of the three bands ($YJH$); two out of 11 are recovered by our search 
criteria: LHS\,2045 (esdM4.5) and SDSS J020533.75$+$123824.0 
\citep[ID=5; esdM8.5;][]{lepine08b}. However, two are not within our sample: 
LHS\,2096 \citep[esdM5.5; Hr=20.62 mag][]{lepine05d,west08} because of the 
lower limit we imposed on the reduced proper motion, and 
SDSS J023557.61$+$010800.5 \citep[esdM7;][]{lepine08b} because of its
separation between the SDSS and UKIDSS coordinates (0.92 arcsec) compared
to our lower limit of 1.0 arcsec.

%
%%%%%%%%%%%%%%%%%%%%%%%%%%%%%%%%%%%%%%%%%%
%%%%%%%%%% Revised PMs %%%%%%%%%
%%%%%%%%%%%%%%%%%%%%%%%%%%%%%%%%%%%%%%%%%%
%

\section{Proper motion revision}
\label{subdw:revised_PM}

In this section we discuss the accuracy of the proper motions obtained
by the Virtual Observatory tools by comparing the positions in the
SDSS and UKIDSS catalogues (no errors considered) and dividing by the
epoch difference.

\subsection{Method}
\label{subdw:revised_PM_method}

\begin{figure}
  \centering
  \includegraphics[width=\linewidth, angle=0]{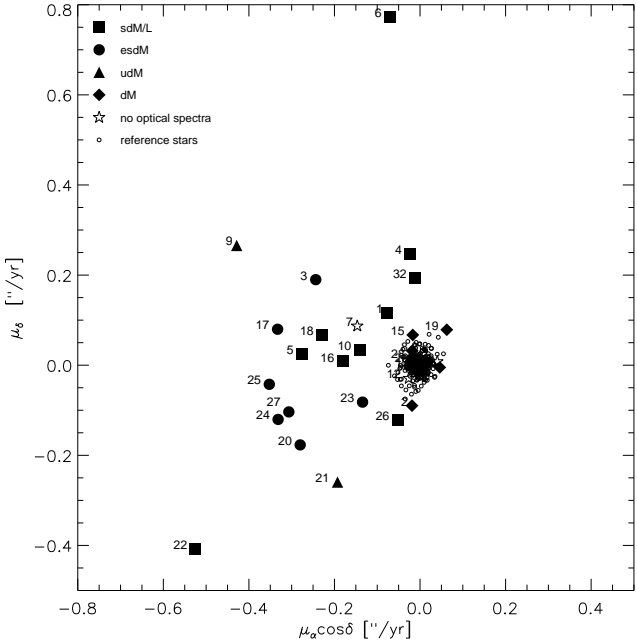}
  \caption{Revised proper motions of our candidates. Astrometric reference 
stars (small open squares) are concentrated around (0,0). Symbols as in 
Fig.\ \ref{fig_subdw:fig_RPM}. Error bars are smaller than the size of
the symbols.
}
  \label{fig_subdw:pmRA_pmDEC}
\end{figure}

We computed accurate proper motions by measuring the pixel coordinates 
($x$,$y$) of the targets and tens of other sources on the Sloan $z$ and 
UKIDSS LAS $J$ images. We carried out this
procedure under the IRAF environment. Firstly, we downloaded the SDSS
and UKIDSS LAS images with a size of 6 arcmin aside, centered on each target.
Secondly, we identify high-quality point sources with a signal-to-noise
higher than 10 in both images, objects selected as reference stars for
proper motion measurements. We assumed that these sources (about 30 per
target) are not moving, assumption valid because they are centered around 
(0,0) in Fig.\ \ref{fig_subdw:pmRA_pmDEC}. 

Then, we transformed the pixel coordinates from one epoch to the other epoch 
using second-order polynomial transformations for both $x$ and $y$ axes. The 
dispersion of the transformations was typically 0.16 pixel (or 0.032 arcsec). 
The resulting $x$ and $y$ shifts were converted into proper motions by 
taking into account the time baseline of the data and the appropriate pixel 
scale values. Our measurements along their associated error bars are given 
in Table \ref{tab_subdw:tab_pmRA_pmDEC_new}.

%
%%%%%%%%%%%%%%%%%%%%%%%%%%%%%%%%%%%%%%%%%%
%%%%% Table: new PMs %%%%%
%%%%%%%%%%%%%%%%%%%%%%%%%%%%%%%%%%%%%%%%%%
%
\begin{table}
 \centering
 \caption[]{Revised proper motions and reduced proper motions along with 
the values derived from the VO cross-match}
 \begin{tabular}{@{\hspace{0mm}}c@{\hspace{1mm}}c@{\hspace{2mm}}c@{\hspace{2mm}}c@{\hspace{2mm}}c@{\hspace{2mm}}c@{\hspace{2mm}}c@{\hspace{0mm}}}
% \begin{tabular}{c c c c c c c}
 \hline
 \hline
ID   & $\mu_{\alpha}cos{\delta}$ & $\mu_{\delta}$ & $\mu_{\rm total}$ & Hr & $\mu$(VO) & Hr(VO)  \\
 \hline
     &  $''$/yr  & $''$/yr  & $''$/yr  &  mag & $''$/yr  &  mag \\
 \hline
   1 & $-$0.077$\pm$0.005 &  0.115$\pm$0.005 & 0.139$\pm$0.007 & 22.215 & 0.137 & 22.177\\
   2 &  0.004$\pm$0.005 & $-$0.103$\pm$0.005 & 0.103$\pm$0.007 & 22.659 & 0.135 & 23.240\\
   3 & $-$0.244$\pm$0.001 &  0.190$\pm$0.002 & 0.309$\pm$0.002 & 21.405 & 0.283 & 21.211\\
   4 & $-$0.022$\pm$0.005 &  0.246$\pm$0.004 & 0.247$\pm$0.006 & 23.050 & 0.253 & 23.106\\
   5 & $-$0.276$\pm$0.002 &  0.024$\pm$0.001 & 0.277$\pm$0.002 & 21.979 & 0.270 & 21.925\\
   6 & $-$0.070$\pm$0.002 &  0.773$\pm$0.001 & 0.776$\pm$0.003 & 26.180 & 0.783 & 26.201\\
   7 & $-$0.147$\pm$0.003 &  0.087$\pm$0.004 & 0.171$\pm$0.005 & 23.662 & 0.166 & 23.592\\
   8 & $-$0.018$\pm$0.006 &  0.032$\pm$0.006 & 0.037$\pm$0.009 & 21.522 & 0.297 & 26.044\\
   9 & $-$0.429$\pm$0.006 &  0.266$\pm$0.005 & 0.504$\pm$0.008 & 21.327 & 0.500 & 22.514\\
  10 & $-$0.141$\pm$0.006 &  0.034$\pm$0.006 & 0.145$\pm$0.008 & 23.127 & 0.162 & 23.364\\
  11 &  0.040$\pm$0.007 &  0.008$\pm$0.008 & 0.041$\pm$0.011 & 20.438 & 0.614 & 26.314\\
  12 & $-$0.026$\pm$0.006 & $-$0.029$\pm$0.006 & 0.039$\pm$0.008 & 20.290 & 1.008 & 27.353\\
  13 &  0.103$\pm$0.007 &  0.069$\pm$0.006 & 0.124$\pm$0.009 & 22.681 & 0.803 & 26.739\\
  14 &  0.046$\pm$0.006 & $-$0.005$\pm$0.006 & 0.046$\pm$0.008 & 21.554 & 0.273 & 25.421\\
  15 & $-$0.017$\pm$0.005 &  0.067$\pm$0.005 & 0.069$\pm$0.007 & 21.707 & 0.162 & 23.556\\
  16 & $-$0.181$\pm$0.003 &  0.008$\pm$0.005 & 0.181$\pm$0.006 & 23.249 & 0.189 & 23.346\\
  17 & $-$0.332$\pm$0.007 &  0.080$\pm$0.007 & 0.341$\pm$0.010 & 23.265 & 0.344 & 23.284\\
  18 & $-$0.229$\pm$0.003 &  0.066$\pm$0.003 & 0.239$\pm$0.005 & 23.409 & 0.255 & 23.545\\
  19 &  0.063$\pm$0.010 &  0.080$\pm$0.010 & 0.101$\pm$0.141 & 22.919 & 0.303 & 25.306\\
  20 & $-$0.280$\pm$0.004 & $-$0.177$\pm$0.003 & 0.331$\pm$0.005 & 22.162 & 0.290 & 21.877\\
  21 & $-$0.193$\pm$0.007 & $-$0.259$\pm$0.008 & 0.323$\pm$0.011 & 22.340 & 0.294 & 22.133\\
  22 & $-$0.525$\pm$0.007 & $-$0.407$\pm$0.007 & 0.664$\pm$0.010 & 26.610 & 0.673 & 26.638\\
  23 & $-$0.134$\pm$0.005 & $-$0.082$\pm$0.004 & 0.157$\pm$0.007 & 21.062 & 0.178 & 21.334\\
  24 & $-$0.332$\pm$0.006 & $-$0.120$\pm$0.006 & 0.353$\pm$0.009 & 22.223 & 0.355 & 22.236\\
  25 & $-$0.352$\pm$0.010 & $-$0.042$\pm$0.012 & 0.355$\pm$0.016 & 23.256 & 0.320 & 23.030\\
  26 & $-$0.052$\pm$0.003 & $-$0.121$\pm$0.004 & 0.132$\pm$0.005 & 22.475 & 0.188 & 23.248\\
  27 & $-$0.307$\pm$0.010 & $-$0.104$\pm$0.010 & 0.324$\pm$0.014 & 21.827 & 0.312 & 21.742\\
  28 & $-$0.016$\pm$0.008 &  0.015$\pm$0.010 & 0.022$\pm$0.013 & 19.136 & 0.326 & 24.992\\
  29 & $-$0.007$\pm$0.006 &  0.008$\pm$0.006 & 0.011$\pm$0.009 & 18.485 & 0.733 & 27.602\\
  30 &   --- $\pm$ ---  &   --- $\pm$ ---  &  --- $\pm$ ---  &  ---   & 1.210 & 28.182\\
  31 &   --- $\pm$ ---  &   --- $\pm$ ---  &  --- $\pm$ ---  &  ---   & 1.129 & 27.983\\
  32 & -0.012$\pm$0.004 &  0.193$\pm$0.005 & 0.193$\pm$0.007 & 22.904 & 0.202 & 23.005\\
 \hline
 \label{tab_subdw:tab_pmRA_pmDEC_new}
 \end{tabular}
\end{table}
%
%
%%%%%%%%%%%%%%%%%%%%%%%%%%%%%%%%%%%%%%
%%%%% (mu_alpha,mu_delta) plot %%%%%
%%%%%%%%%%%%%%%%%%%%%%%%%%%%%%%%%%%%%%
%
\subsection{New proper motions}
\label{subdw:revised_PM_results}

We find that the revised proper motions agree with the proper motion obtained
from the VO for 22 of the 32 candidates ($\sim$70\%) 
(Table \ref{tab_subdw:tab_pmRA_pmDEC_new})
Below we give details on the 10 sources whose new proper motion differ
from the original ones, leading to a different position in the
reduced proper motion diagram (Table \ref{tab_subdw:tab_pmRA_pmDEC_new}; 
Fig.\ \ref{fig_subdw:fig_RPM}):
\begin{itemize}
\item Sources \#30 and \#31 should be rejected because no object is detected
on the UKIDSS images suggesting that these are false cross-matches. No optical
spectra (Section \ref{subdw:spec_obs}) are available for these sources so we 
reject them from our sample
\item Sources \#11 and \#12 have smaller proper motion because of large offsets
between the position of the source on the SDSS image and the position reported
by the SDSS catalogue. Their revised H$r$ is smaller than 20.7 mag, implying
that they would not enter our sample. No optical spectra are available for
them so we reject them
\item Three sources classified as M dwarfs (\#8, 28, and 29) are rejected
because their revised proper motions are smaller than the one derived by the
VO, leading to smaller H$r$ value and/or a position in the reduced proper
motion diagram suggesting that they are solar-metallicity dwarfs
\item Three M dwarfs (\#14, 15, and 19) have smaller revised proper motions
but lie in the same region as subdwarfs in the reduced proper motion diagram
\end{itemize}

We note that sources \#7 and \#2 have revised proper motions identical to 
those from the VO and both objects lie in the subdwarf domain in the reduced 
proper motion diagram. Source \#7 has no optical spectroscopy, it thus 
remains a reliable ultracool subdwarf candidate. On the contrary, source \#2 
has optical spectroscopy suggesting it is a solar-metallicity M dwarf 
contaminating our sample.

To summarise, we conclude that only one of the five sources without optical
spectrum remains as a bona-fide subdwarf. Moreover, three of the seven sources
spectroscopically classified as M dwarfs (Fig.\ \ref{fig_subdw:fig_dM}) 
would be rejected
based on the revised proper motion, the remaining ones being photometric
contaminants because they lie in the region of the reduced proper motion
diagram where ultracool subdwarfs are found (right panel in
Fig.\ \ref{fig_subdw:fig_RPM}).

%
%%%%%%%%%%%%%%%%%%%%%%%%%%%%%%%%%%%%%%%%%%
%%%%%%%%%% Observations %%%%%%%%%
%%%%%%%%%%%%%%%%%%%%%%%%%%%%%%%%%%%%%%%%%%
%
\section{Optical spectroscopy}
\label{subdw:spec_obs}

We emphasise that optical spectroscopy was obtained for candidates 
identified in the preliminary search i.e.\ with the proper motion derived 
from the VO and not from the revised proper motions discussed in 
Section \ref{subdw:revised_PM}.

\subsection{NOT spectroscopy}
\label{subdw:obs_NOT}

We carried out low-resolution (R$\sim$450) optical (500--1025 nm)
spectroscopy with the ALFOSC spectrograph on the 2.5-m Nordic Optical Telescope 
(NOT) at the Observatory of the Roque de Los Muchachos on the island of La
Palma, Canary Islands. One candidate, ULAS J145441.41$+$123556.6 (ID=27), was 
observed on 3 September 2011 (Table \ref{tab_subdw:tab_log_spectro}). Weather 
conditions were non photometric with 
high level clouds and seeing around 0.7--0.9 arcsec. The object was observed 
around UT = 21h at parallactic angle under high airmass, starting at 1.5 and 
finishing around 1.9 (Table \ref{tab_subdw:tab_log_spectro}).

The ALFOSC spectrograph is equipped with a 2048\,$\times$\,2052 pixel 
back-illuminated CCD42-40 charge coupled device. We employed the grism 
number 5\@. The total exposure time for ULAS J145441.41$+$123556.6 was divided 
into three on-source integrations of 800 seconds with a slit width of 1 arcsec 
to achieve a resolution of 1.55 nm at 700 nm.
An internal flat-field was obtained immediately after the target to remove 
as much as possible the fringing at long wavelength, the effect being of the 
order of 18\% at 800 nm. Unfortunately fringing is clearly visible
beyond 800 nm because the internal flat field was obtained after the last of
the three individual exposures taken for the target. Bias frames were observed 
in the afternoon before the beginning of the night. He, Ne, and Ar arc lamps 
were also obtained immediately after each exposure to calibrate our target in 
wavelength.

The data reduction of the NOT/ALFOSC spectra was entirely carried out under the 
IRAF\footnote{IRAF is distributed by the National Optical Astronomy Observatory, 
which is operated by the Association of Universities for Research in Astronomy 
(AURA) under cooperative agreement with the National Science Foundation} 
environment. First, we subtracted the average of all bias frames from the raw 
science exposures and then divided by the normalised response function of the 
mean flat field (also bias subtracted). Second, we extracted a one-dimensional 
spectrum interactively by choosing the appropriate width for the aperture width.
Then, we used the arc lamps to calibrate our spectra in wavelength with
an accuracy better than 0.2\,\AA{}. The flux calibration of the spectrum was 
conducted using BD$+$174708 \citep{latham02} as spectrophotometric standard 
star. The final optical spectrum, covering the 500--900 nm wavelength range 
and normalised at 750 nm is shown in Fig.\ \ref{fig_subdw:fig_esdM}.

\subsection{FORS2 spectroscopy}
\label{subdw:obs_FORS2}

We conducted low-resolution (R$\sim$350) optical (600--1010 nm) spectroscopy 
with the visual and near UV FOcal Reducer and low dispersion Spectrograph FORS2 
\citep{appenzeller98} on the ESO VLT Antu unit in Paranal, Chile. We observed 
22 subdwarf candidates identified in the UKIDSS LAS DR5 vs SDSS DR7 
cross-match
(Table \ref{tab_subdw:tab_log_spectro}). All observations were conducted in 
service mode under programs 084.C-0928A and 084.C-0928B\@. The requested 
conditions, grey time, thin cirrus acceptable, and seeing less than 1.4 and 
1.0 arcsec for the bright and faint targets, respectively, were generally 
satisfied (Table \ref{tab_subdw:tab_log_spectro}). All objects were observed
at parallactic angle.

The FORS2 instrument is equipped with two 2048$\times$4096 MIT CCDs with 
pixels of 15$\mu$m working in the 330--1100 nm range \citep{appenzeller98}. 
We employed the grism 150$+$27 with the order blocking filter OG590 with the 
standard resolution of 2.07 nm per pixel and a slit of 1.0 arcsec to achieve 
a spectral resolution of $\sim$175 at 720 nm due to the 2$\times$2 binning.
The exposure time was scaled according to the brightness of the target in the 
Sloan $r$ and $i$ filter sin order to achieve a minimum signal-to-noise of 20\@.
The faintest sources, observed several times along the slit, have lower 
signal-to-noise because they are faint even for a 8-m class telescope.
Dome flat fields, bias frames, arc lamps, and spectrophotometric standard 
stars were observed every night as part of the ESO calibration plan.

\begin{figure}
  \centering
  \includegraphics[width=\linewidth, angle=0]{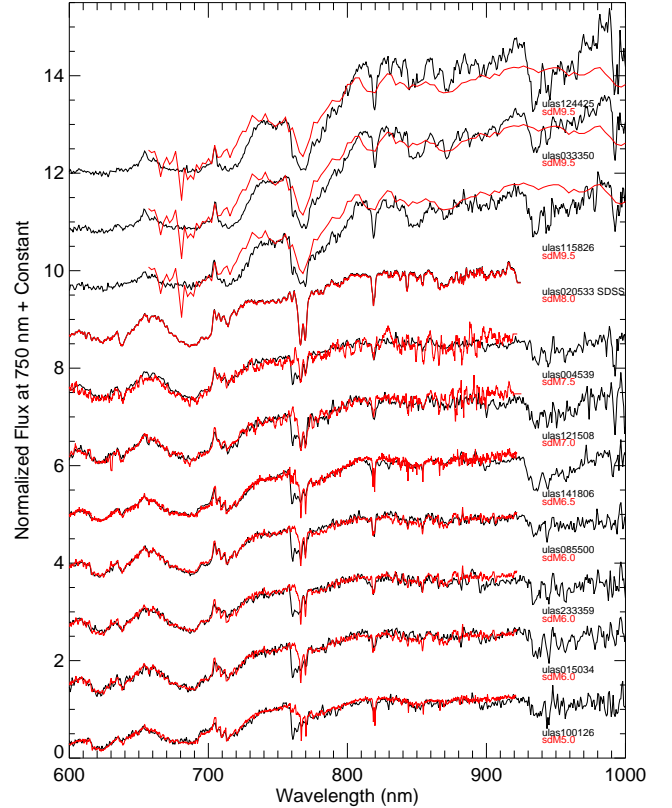}
  \caption{
Low-resolution optical (600-1000 nm) spectra of confirmed subdwarfs
obtained with VLT FORS2\@. Spectra are ordered by increasing spectral type
and offset by a constant value for clarity. Overplotted in red are known 
template subdwarfs from the SDSS spectroscopic database (600--940 nm)
except for the top 3 templates which come from the IRTF/SpeX library.
}
  \label{fig_subdw:fig_sdM}
\end{figure}

\begin{figure}
  \centering
  \includegraphics[width=\linewidth, angle=0]{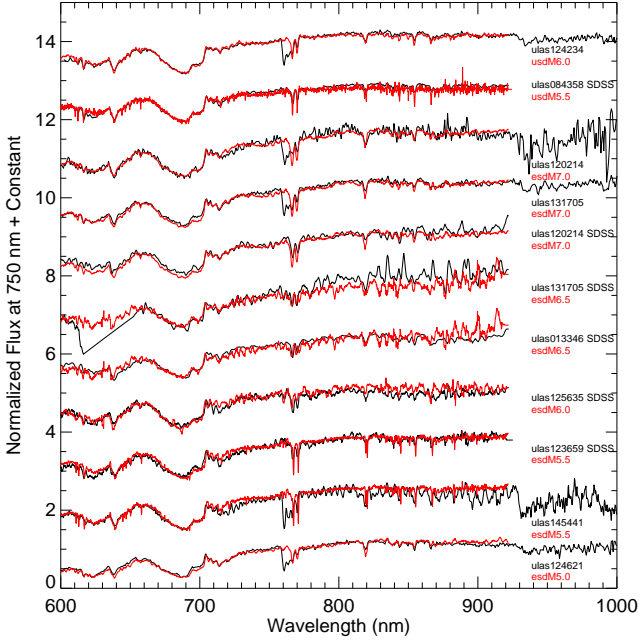}
  \caption{
Low-resolution optical (600-1000 nm) spectra of confirmed extreme 
subdwarfs and ultra-subdwarfs obtained with VLT FORS2\@. The second
spectrum from bottom is from NOT/ALFOSC\@. Spectra are ordered by increasing 
spectral type and offset by a constant value for clarity. Overplotted in red 
are known template extreme subdwarfs and ultra-subdwarfs from the SDSS 
spectroscopic database (600--940 nm).
}
  \label{fig_subdw:fig_esdM}
\end{figure}

\begin{figure}
  \centering
  \includegraphics[width=\linewidth, angle=0]{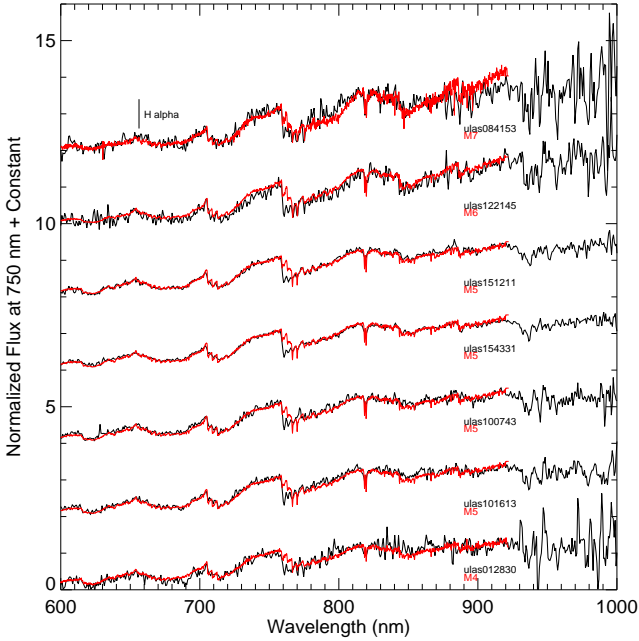}
  \caption{
Low-resolution optical (600-1000 nm) spectra of photometric candidates
classified as solar-metallicity M dwarfs. Spectra are ordered by increasing 
spectral type and offset by a constant value for clarity. Overplotted in red 
are known M dwarf templates downloaded from the SDSS spectroscopic database.
}
  \label{fig_subdw:fig_dM}
\end{figure}

As for NOT/ALFOSC, the data reduction of the VLT FORS2 spectra was entirely 
carried out under the IRAF environment. First, we cut the 2D images to select 
the interesting part of the spectrum, from $\sim$600 to 1000 nm roughly. Then, 
we subtracted the average of all bias frames from the raw science exposures 
and from the median-combined dome flat. Afterwards, we divided the science
frame by the normalised flat field using the mean value over the entire
dome flat frame. Later, we extracted a one-dimensional spectrum
interactively by choosing the appropriate value for the aperture width.
Then, we used the arc lamps with Helium, HgCd, and Argon to calibrate our 
spectra in wavelength with an accuracy of the order of 0.4--0.6\,\AA{} rms. 
Finally we calibrated the 1D spectra with spectrophotometric standard stars 
observed on the same night as the target. Spectra have been normalised at 
750 nm and are not corrected for the telluric band around 760 nm. 
The VLT FORS2 spectra of the 11 candidates classified as subdwarfs are 
displayed in Fig.\ \ref{fig_subdw:fig_sdM}. The sources classified as 
extreme subdwarfs (7 objects) and ultra-subdwarf (2 objects) are shown in 
Fig.\ \ref{fig_subdw:fig_esdM} while the seven sources classified as 
solar-metallicity M dwarfs are displayed in Fig.\ \ref{fig_subdw:fig_dM}. 
The two known subdwarfs previously reported in the literature are included 
in these figures.

%
%%%%%%%%%%%%%%%%%%%%%%%%%%%%%%%%%%%%%%%%%%
%%%%% Table Log spectroscopy %%%%%
%%%%%%%%%%%%%%%%%%%%%%%%%%%%%%%%%%%%%%%%%%
%
\begin{table}
 \centering
 \caption[]{Log of the spectroscopic observations}
 \begin{tabular}{@{\hspace{0mm}}c@{\hspace{2mm}}c@{\hspace{2mm}}c@{\hspace{2mm}}c@{\hspace{2mm}}c@{\hspace{2mm}}c@{\hspace{2mm}}c@{\hspace{2mm}}c@{\hspace{0mm}}}
 \hline
 \hline
R.A.        &  Dec          & Instr & Date           & \# & ExpT & Airm & Seeing \cr
 \hline
hh:mm:ss.ss & $^{\circ}$:$'$:$''$ &   & yyyy-mm-dd     &    &  sec  &     &  $''$  \cr
 \hline
00:45:39.97 & $+$13:50:32.7 & FORS  & 2009-11-10     &  3 &  1170 & 1.286 & 0.74 \cr
01:28:30.89 & $+$13:45:07.4 & FORS  & 2009-11-04/15  &  4 &  2640 & 1.286 & 0.62 \cr
01:50:34.33 & $+$14:20:02.4 & FORS  & 2009-10-30     &  3 &  1170 & 1.494 & 0.59 \cr
03:33:50.84 & $+$00:14:06.1 & FORS  & 2009-10-30     &  3 &   870 & 1.546 & 0.80 \cr
08:41:53.89 & $+$02:06:15.1 & FORS  & 2009-12-21/25  &  4 &  2640 & 1.121 & 0.81 \cr
08:55:00.12 & $+$00:02:04.1 & FORS  & 2009-12-25     &  1 &   660 & 1.122 & 0.65 \cr
10:01:26.29 & $-$01:34:26.6 & FORS  & 2010-01-08     &  2 &  1330 & 1.092 & 0.86\cr
10:07:43.96 & $-$02:28:30.0 & FORS  & 2009-12-17/24  &  4 &  2644 & 1.150 & 0.83 \cr
10:16:13.89 & $+$01:13:11.4 & FORS  & 2009-12-24     &  2 &  1320 & 1.161 & 0.56 \cr
11:58:26.62 & $+$04:47:46.8 & FORS  & 2009-12-21     &  1 &  1320 & 1.468 & 0.73 \cr
12:02:14.62 & $+$07:31:13.8 & FORS  & 2010-01-23     &  1 &   390 & 1.222 & 0.77 \cr
12:15:08.37 & $+$04:02:00.5 & FORS  & 2010-02-12     &  3 &  1170 & 1.183 & 0.87 \cr
12:21:45.28 & $+$08:04:04.4 & FORS  & 2010-01-08/09  &  4 &  2644 & 1.280 & 0.60 \cr
12:42:34.62 & $+$14:33:06.2 & FORS  & 2010-03-07     &  1 &   390 & 1.306 & 0.53 \cr
12:44:25.90 & $+$10:24:41.9 & FORS  & 2010-01-23     &  1 &   390 & 1.237 & 0.95 \cr
12:46:21.90 & $+$04:43:09.9 & FORS  & 2010-03-07     &  1 &   390 & 1.181 & 0.57 \cr
13:17:05.66 & $+$09:10:16.9 & FORS  & 2010-03-07     &  1 &   390 & 1.239 & 0.52 \cr
14:18:06.71 & $+$00:00:35.5 & FORS  & 2010-02-13     &  3 &  1170 & 1.119 & 0.71 \cr
15:12:11.64 & $+$06:42:51.3 & FORS  & 2010-03-07     &  4 &  2640 & 1.510 & 0.57 \cr
15:43:31.93 & $+$02:45:37.8 & FORS  & 2010-03-07     &  8 &  5280 & 1.333 & 0.63 \cr
23:33:59.39 & $+$00:49:35.2 & FORS  & 2009-11-10     &  3 &  1170 & 1.130 & 1.14 \cr
 \hline
14:54:41.42 & $+$12:35:56.7 & NOT   & 2011-09-03     &  3 &  2400 & 1.700 & 0.80 \cr
 \hline
01:33:46.25 & $+$13:28:22.4 & SDSS  & 2000-12-01     &  5 &  5400 & 1.143 & 2.82 \cr
02:05:33.75 & $+$12:38:24.1 & SDSS  & 2000-12-01     &  5 &  4500 & 1.143 & 2.82 \cr
08:43:58.50 & $+$06:00:38.6 & SDSS  & 2003-03-10     &  3 &  2400 & 1.161 & 1.81 \cr
12:02:14.62 & $+$07:31:13.8 & SDSS  & 2004-03-25     &  3 &  3000 & 1.133 & 1.41 \cr
12:36:59.43 & $-$00:21:58.2 & SDSS  & 2001-02-01     &  4 &  3600 & 1.209 & 2.87 \cr
12:56:35.91 & $-$00:19:44.9 & SDSS  & 2001-03-26     &  3 &  2702 & 1.214 & 2.28 \cr
13:17:05.66 & $+$09:10:16.9 & SDSS  & 2006-04-25     &  4 &  4900 & 1.216 & 2.02 \cr
 \hline
 \label{tab_subdw:tab_log_spectro}
 \end{tabular}
\end{table}
\subsection{SDSS spectroscopy}
\label{subdw:obs_SDSS}

The SDSS spectroscopic database represents an invaluable source of good quality
spectra for a large number of astronomical sources. These spectra, covering
the 380--940 nm wavelength range with a resolution of $\sim$2000 are publicly 
available from the Sloan webpage. They are wavelength and flux calibrated
and also corrected for telluric absorption bands. The total exposure times
consist of several integrations of 900 seconds. We found seven of our 32 
ultracool subdwarf candidates in this database, two of them 
(ULAS J131705.66$+$091016.9 (ID=25) and ULAS J120214.62$+$073113.8 (ID=17)) 
re-observed with VLT 
FORS2 because of their poorer quality (Table \ref{tab_subdw:tab_log_spectro}). 
The other five SDSS spectra have good signal-to-noise and we analysed them in 
a similar manner as the VLT FORS2 spectra (see Sect.\ \ref{subdw:subdw}).
The observing date, number of exposures, airmass and best 80\% seeing are
listed at the bottom of Table \ref{tab_subdw:tab_log_spectro}.

%
%%%%%%%%%%%%%%%%%%%%%%%%%%%%%%%
%%%%% Figure: sdM & sdL %%%%%
%%%%%%%%%%%%%%%%%%%%%%%%%%%%%%%
%
%

%
%%%%%%%%%%%%%%%%%%%%%%%%%%%%%%%
%%%%% Figure: esdM %%%%%
%%%%%%%%%%%%%%%%%%%%%%%%%%%%%%%
%
%

%
%%%%%%%%%%%%%%%%%%%%%%%%%%%%%%%
%%%%% Figure: dM %%%%%
%%%%%%%%%%%%%%%%%%%%%%%%%%%%%%%
%
%

%
%%%%%%%%%%%%%%%%%%%%%%%%%%%%%%%
%%%%% SUBDWARFS %%%%%
%%%%%%%%%%%%%%%%%%%%%%%%%%%%%%%
%
\section{New ultracool subdwarfs}
\label{subdw:subdw}

In this section, we assign spectral types to the new subdwarfs, measure their 
radial velocities, estimate their spectroscopic distances, discuss their 
mid-infrared properties, search for wide companions, and present a preliminary
estimate of their surface density.

%
%%%%%%%% Spectral type %%%%%%%%%%%
%
\subsection{Spectral types}
\label{subdw:spec_SpT}

We employed two independent but complementary methods to assign spectral
types to our new ultracool subdwarfs.

The adopted classication for M-type subdwarfs and extreme subdwarfs 
relies on the scheme proposed by Gizis (1997) and extends to spectral 
types sdM7 and esdM5.5, respectively. This scheme is based 
on the strength of CaH (temperature index) and 
TiO (metallicity index) bands. An extension to later spectral
types (up to sdM9) has been more recently proposed by \citet{lepine03b}.

\begin{figure*}
  \centering
  \includegraphics[width=0.48\linewidth, angle=0]{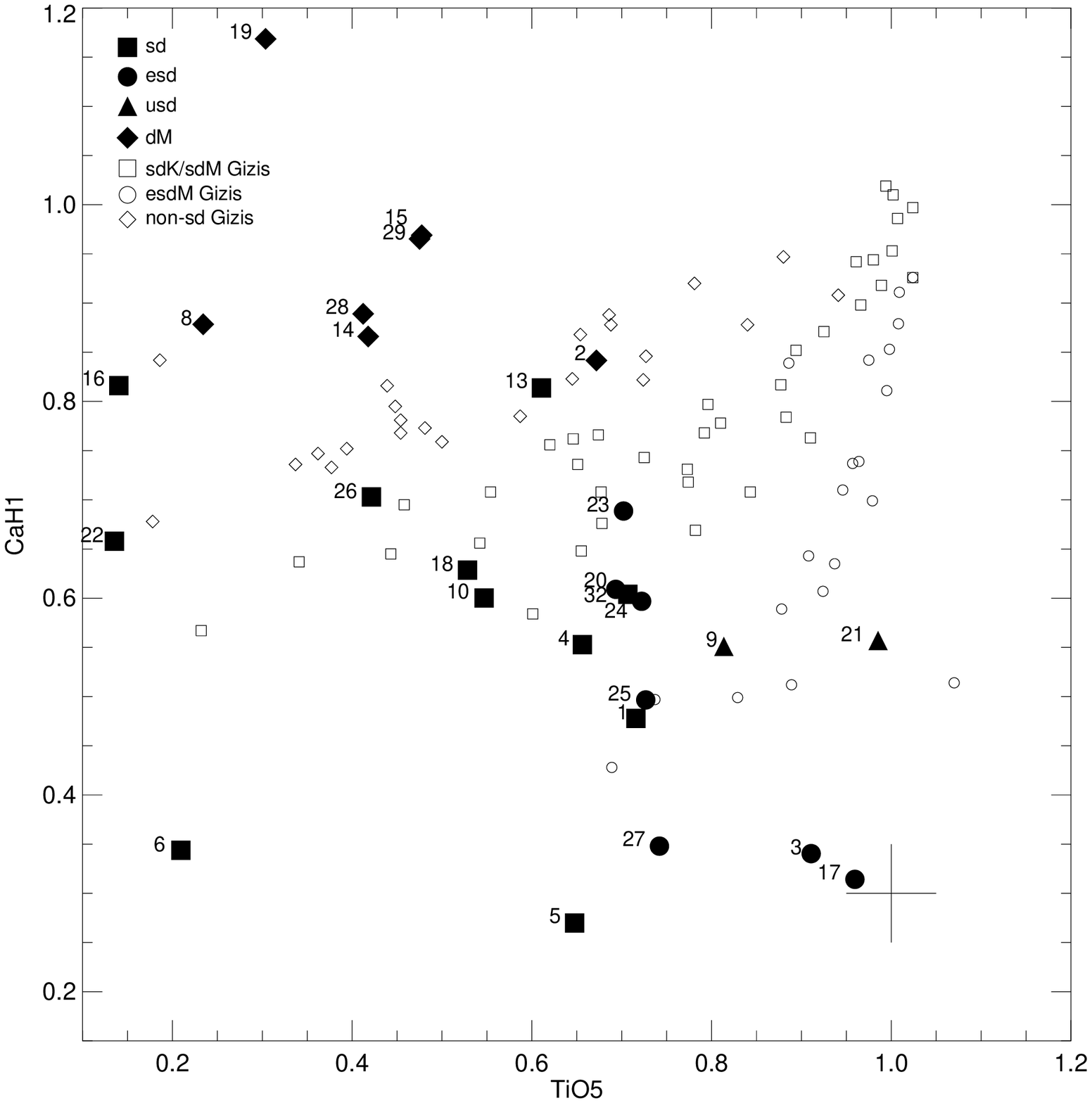}
  \includegraphics[width=0.48\linewidth, angle=0]{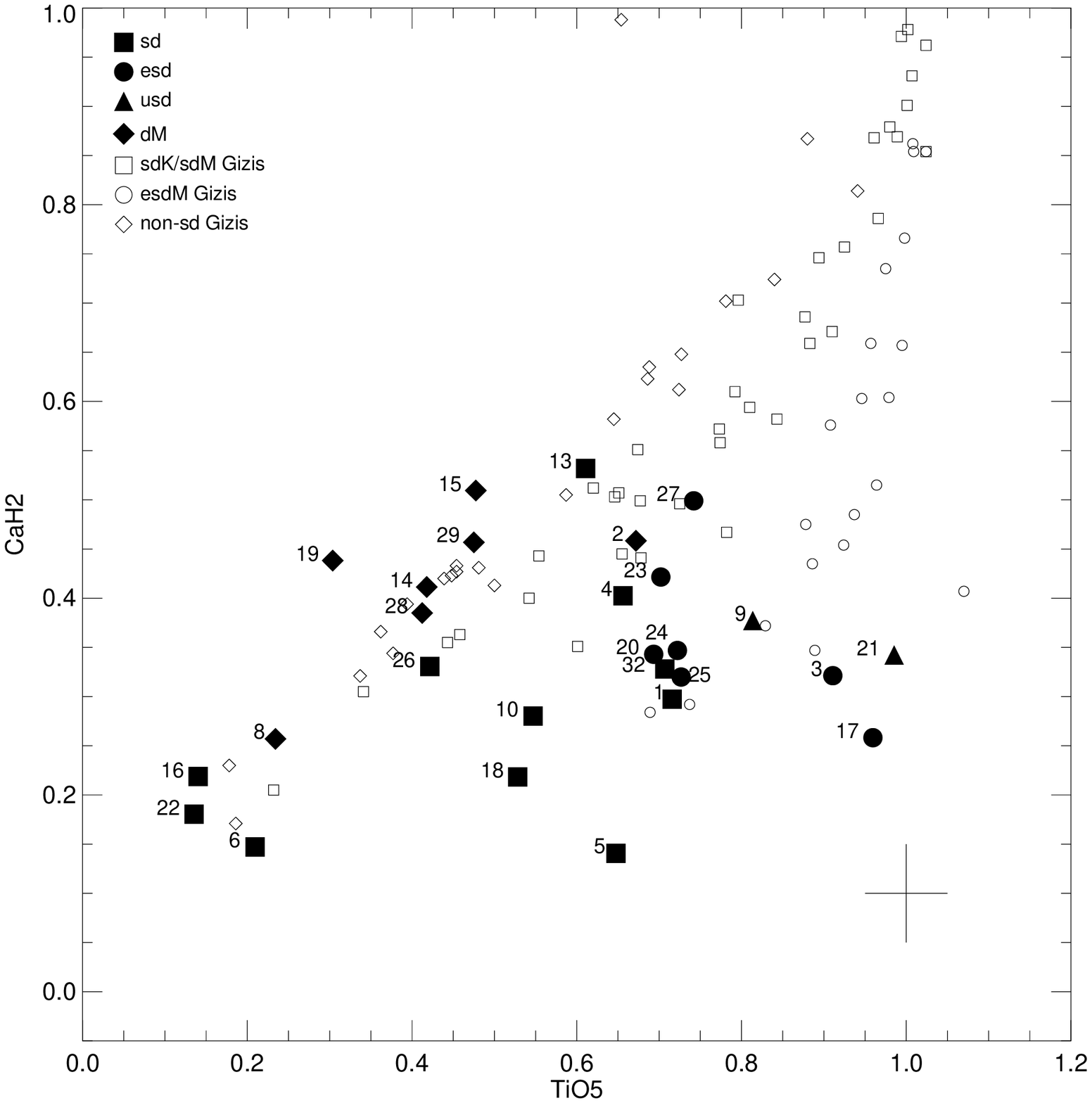}
  \includegraphics[width=0.48\linewidth, angle=0]{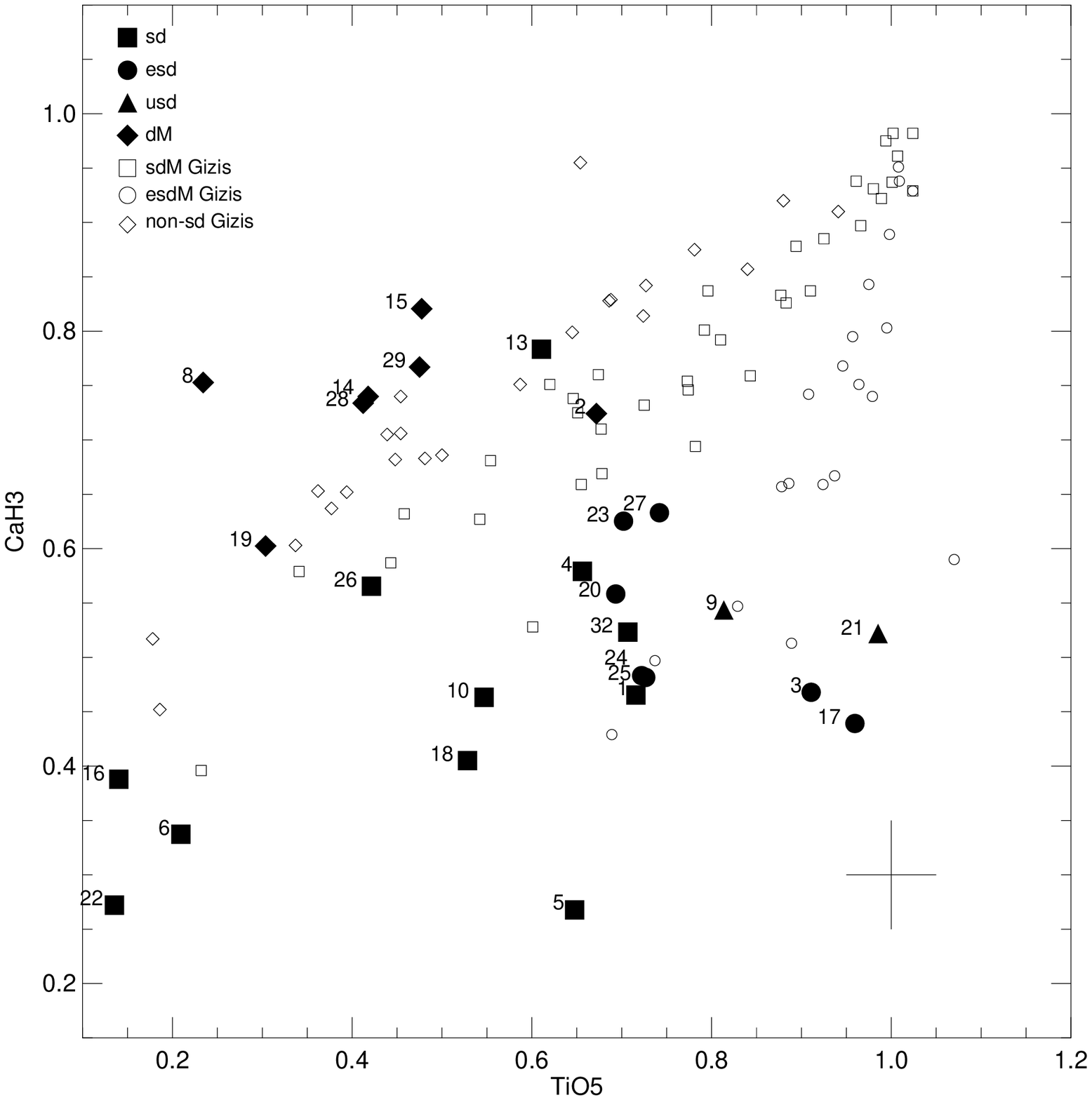}
  \includegraphics[width=0.48\linewidth, angle=0]{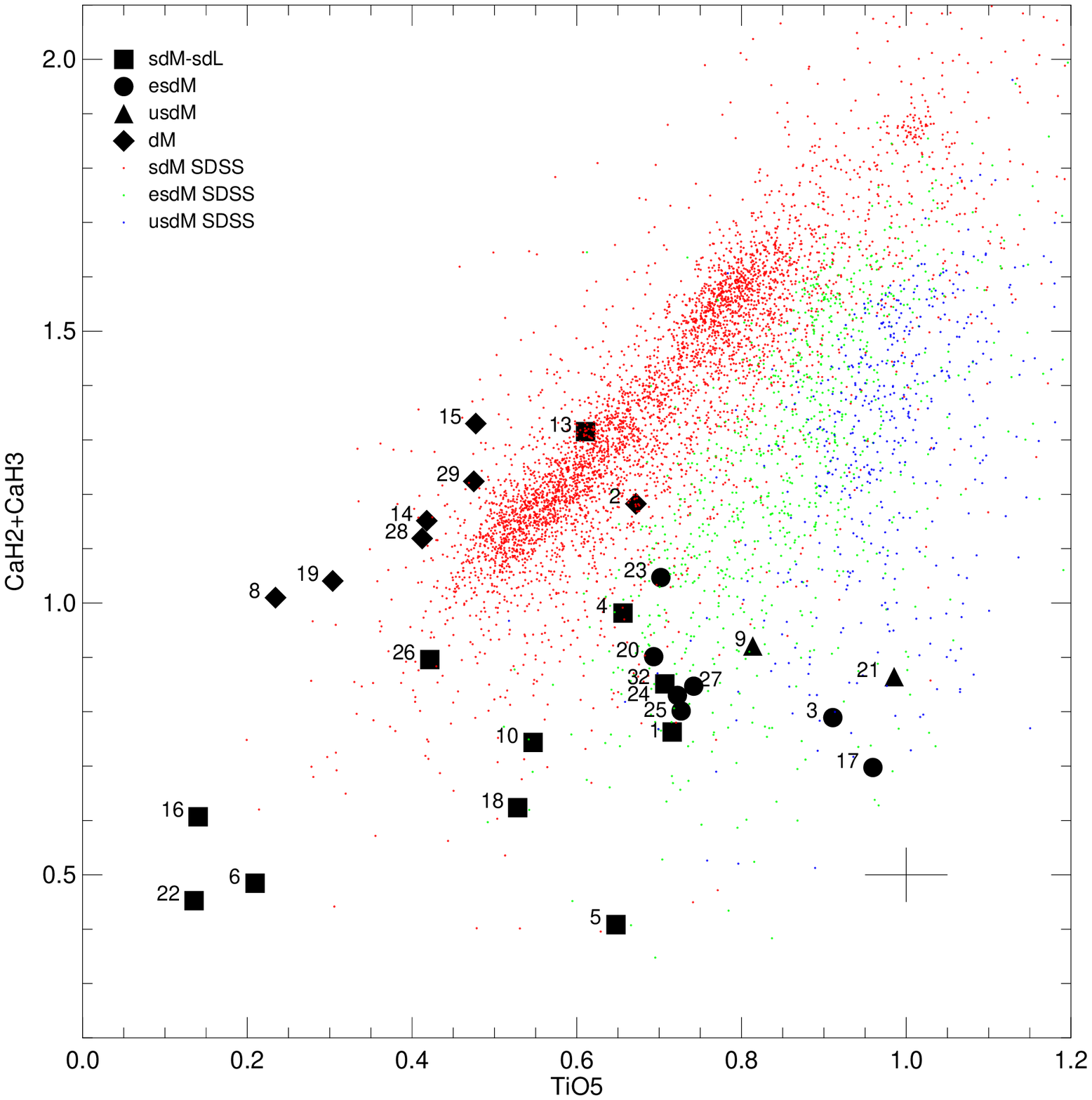}
  \caption{
{\it{Top left:}} CaH1 vs TiO5 indices for our new ultracool subdwarfs.
{\it{Top right:}} CaH2 vs TiO5 indices.
{\it{Bottom left:}} CaH3 vs TiO5 indices.
{\it{Bottom right:}} Sum of CaH2 and CaH3 vs TiO5 indices.
Subdwarfs, extreme subdwarfs, ultra-subdwarfs, old solar-metallicity M dwarfs from 
our sample are marked as filled squares, circles, triangles, and diamonds, 
respectively. Numbers denominating our new discoveries follow the order by right 
ascension from Table \ref{tab_subdw:tab_SpT_indices}. Open symbols are known 
subdwarfs discussed in \citet{gizis97a}. The small coloured dots in the bottom 
right plot represent sources with SDSS spectroscopy classified as subdwarfs (red), 
extreme subdwarfs (green), and ultra-subdwarfs (blue). For the subdwarfs with 
two spectra from FORS2 and SDSS, we plotted only the indices derived from the 
VLT FORS2 spectra. Typical uncertainties on the spectral indices are of the order
of 0.1 (cross at the bottom of each plot).
These plots follow the standard figures presented in Figure 1 
of \citet{gizis97a} and Figure 3 of \citet{lepine07c}.
}
  \label{fig_subdw:fig_SpT_indices}
\end{figure*}

\begin{figure*}
  \centering
  \includegraphics[width=0.48\linewidth, angle=0]{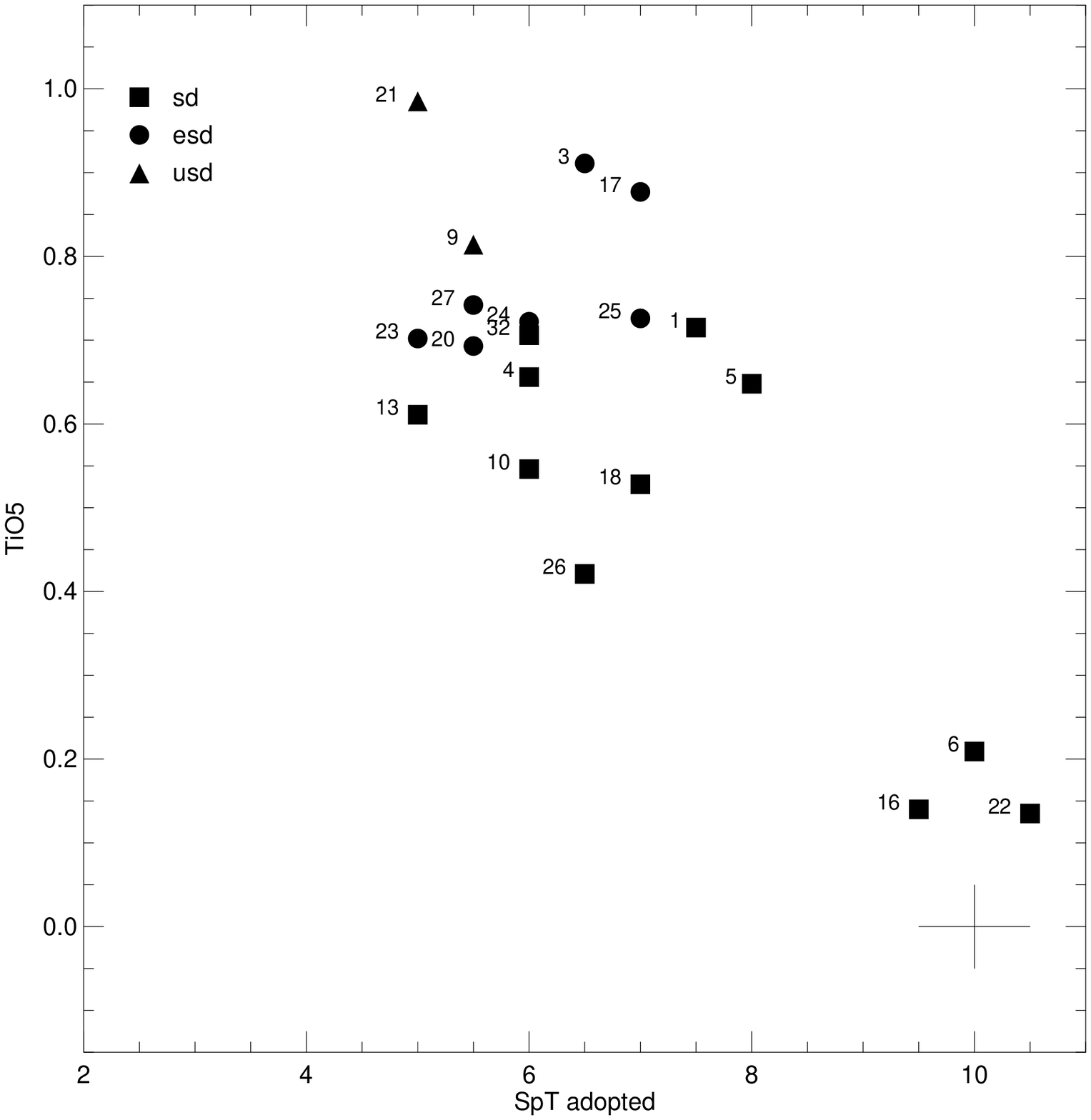}
  \includegraphics[width=0.48\linewidth, angle=0]{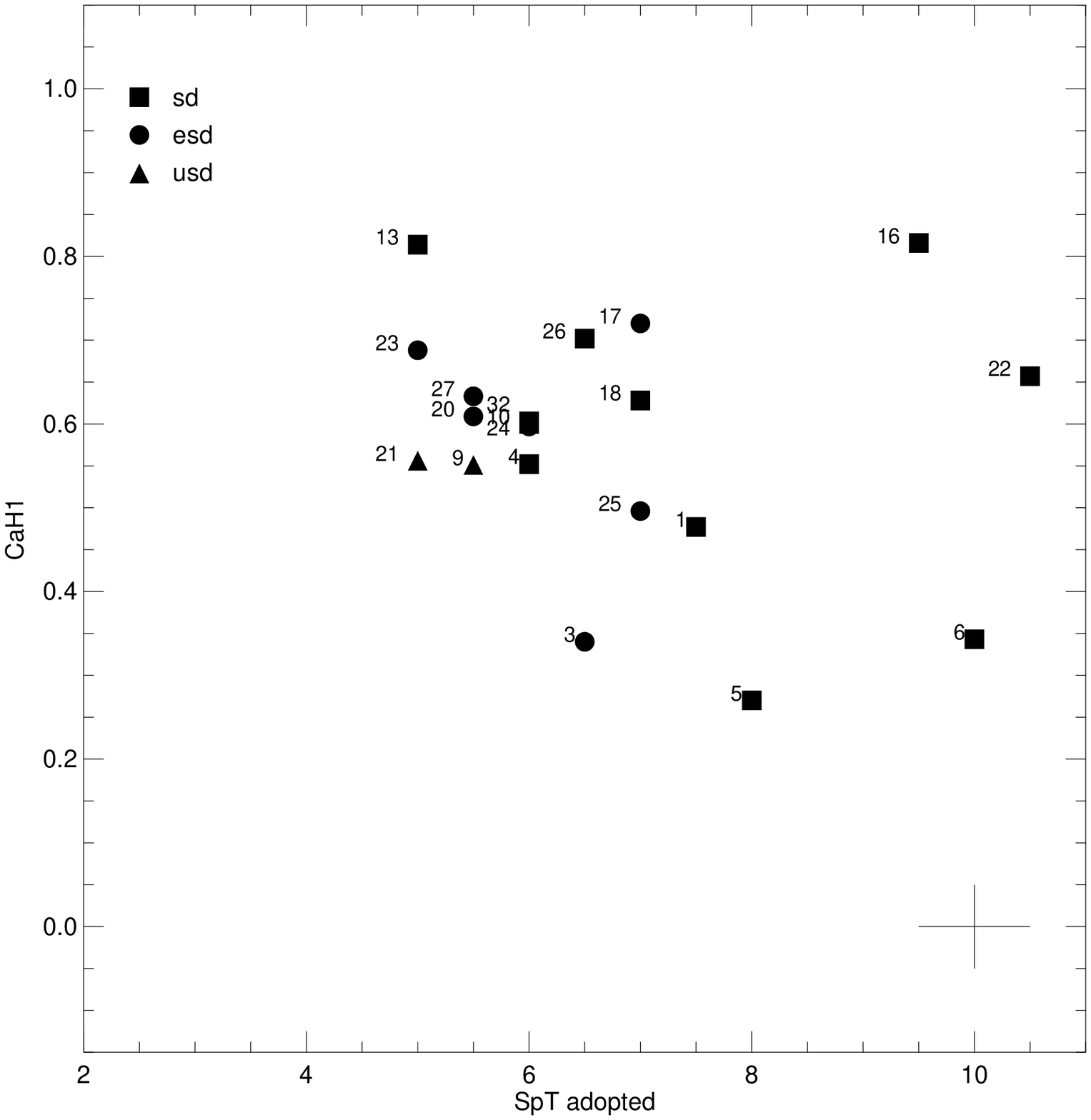}
  \includegraphics[width=0.48\linewidth, angle=0]{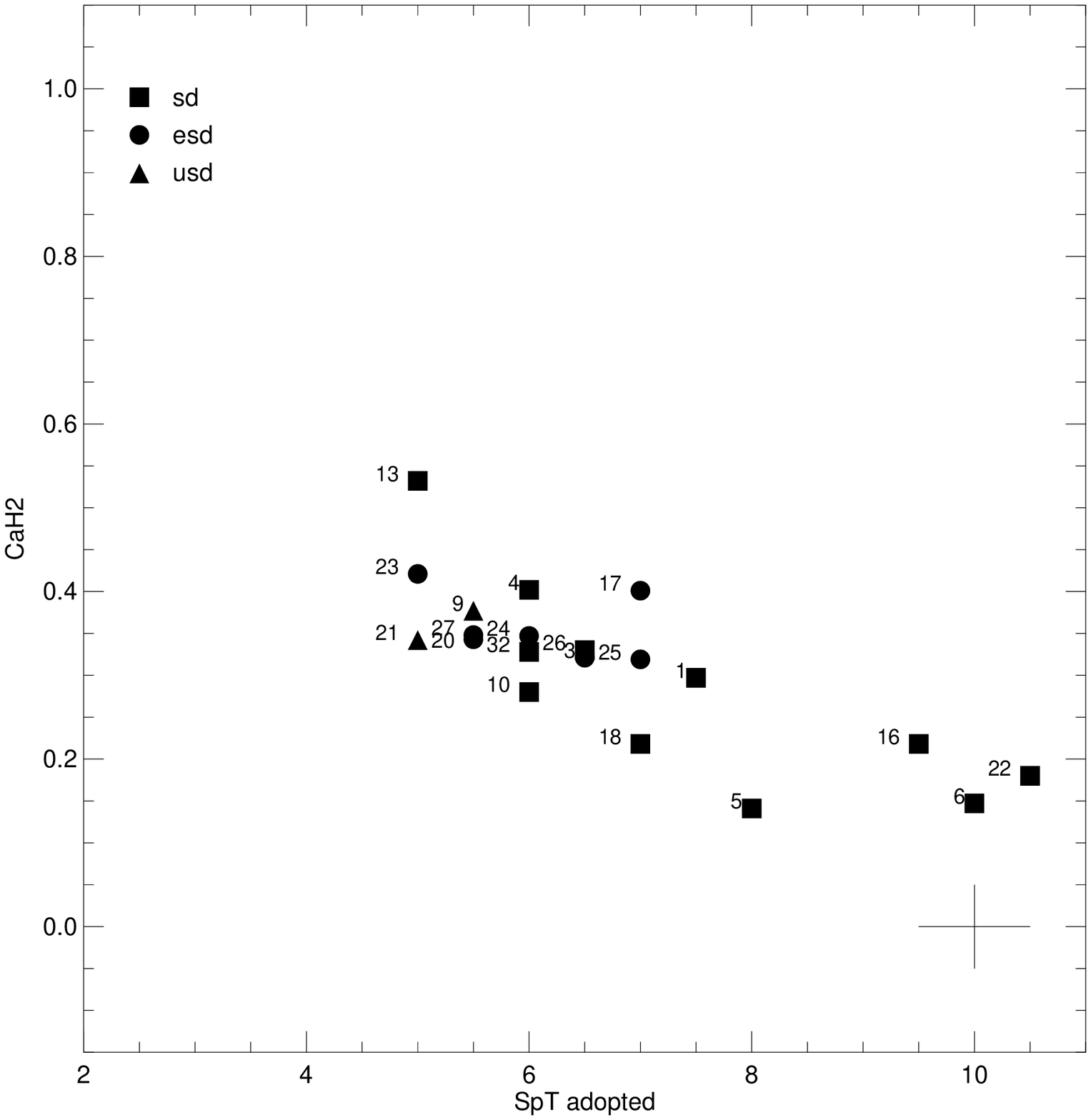}
  \includegraphics[width=0.48\linewidth, angle=0]{18717figC.ps}
  \caption{
Spectral indices as a function of the adopted spectral types for our new
ultracool subdwarfs. Typical uncertainties on the spectral indices are of the 
order of 0.1 (cross at the bottom of each plot). Uncertainties on the
spectral types are 0.5 subclass.
{\it{Top left:}} Adopted spectral types vs TiO5 index.
{\it{Top right:}} Adopted spectral types vs CaH1 index.
{\it{Bottom left:}} Adopted spectral types vs CaH2 index.
{\it{Bottom right:}} Adopted spectral types vs CaH3 index.
}
  \label{fig_subdw:fig_SpT_indices2}
\end{figure*}

We measured the four spectral indices (TiO5, CaH1, CaH2, and CaH3) defined by 
\citet{gizis97a} and later updated by \citet{lepine07c} to distinguish 
ultra-subdwarfs, extreme subdwarfs, and subdwarfs from solar metallicity 
M dwarfs based on the strength of the CaH and TiO absorption bands
(Fig.\ \ref{fig_subdw:fig_SpT_indices}). The spectral 
indices and their associated spectral types (quoted to the nearest decimal)
derived for each of the confirmed subdwarfs are listed in 
Table \ref{tab_subdw:tab_SpT_indices} and plotted in 
Fig.\ \ref{fig_subdw:fig_SpT_indices}--\ref{fig_subdw:fig_SpT_indices2}.

The SDSS spectroscopic database provides optical spectra over the 
600--940 nm wavelength range for a large number of metal-poor dwarfs 
with spectral types based on two independent classification 
schemes: the Hammer scheme discussed by \citet{covey07} and the updated 
subdwarf classification proposed by \citet{lepine07c}. The former distinguishes
between solar-metallicity M dwarfs and subdwarfs whereas the latter provides 
an accurate classification for metal-poor dwarfs with subclasses. Therefore, 
we downloaded from the Sloan spectroscopic database the brightest object of each 
subclass between 0 and 9 for subdwarfs, extreme subdwarfs, and ultra-subdwarfs. 
These ``templates'' were used to assign visually spectral types to our targets
with an uncertainty of half a subclass (or better). We should mention 
that the SDSS templates have been corrected for telluric absorption whereas
our spectra were not, resulting in a difference between them around the position 
of the O$_{2}$ telluric line around 760 nm. We note that the sdM9.5 
template was taken from the SpeX 
library\footnote{http://web.mit.edu/ajb/www/browndwarfs/spexprism/}
because no sdM9.5 template was found in the SDSS spectroscopic database.
The resulting fits (red) to our spectra (black) are presented in 
Figures \ref{fig_subdw:fig_sdM} to \ref{fig_subdw:fig_dM}.

%
%%%%%%%%%%%%%%%%%%%%%%%%%%%%%%%%%%%%%%%%%%
%%%%% Table Spectral Indices %%%%%
%%%%%%%%%%%%%%%%%%%%%%%%%%%%%%%%%%%%%%%%%%
%
\begin{table*}
 \centering
 \caption[]{Coordinates (in J2000), spectral indices, spectral types determined
           following the definitions by \citet{gizis97a} and \citet{lepine07c}
           for the new subdwarfs.
           The last column gives the adopted spectral types derived from direct 
           comparison with spectral templates. If a target appears twice, the
           first line corresponds to the FORS2 spectrum while the second is the 
           SDSS spectrum.}
 \begin{tabular}{c c c c c c c c c c c}
 \hline
 \hline
ID & R.A.        &  Dec          & TiO5  & CaH1  &  CaH2 &  CaH3 &  TiO5 &  SpT\_Gizis & SpT\_Lepine &  SpT\_final \cr
 \hline
   & hh:mm:ss.ss & $^{\circ}$:$'$:$''$ &   &       &       &       &       &             &             &             \cr 
 \hline
 1 & 00:45:39.97 & $+$13:50:32.7 & 0.715 & 0.477 & 0.297 & 0.465 & 0.176 & esdM5.2 & esdM5.6 &  sdM7.5  \cr
 3 & 01:33:46.25 & $+$13:28:22.4 & 0.911 & 0.340 & 0.321 & 0.468 & 0.194 & esdM5.2 & usdM5.4 & esdM6.5  \cr
 4 & 01:50:34.33 & $+$14:20:02.4 & 0.656 & 0.552 & 0.402 & 0.579 & 0.324 &  sdM4.4 &  sdM3.9 &  sdM6.0  \cr
 5 & 02:05:33.75 & $+$12:38:24.1 & 0.648 & 0.270 & 0.141 & 0.268 & 0.002 & esdM7.9 & esdM8.5 &  sdM8.0  \cr
 6 & 03:33:50.84 & $+$00:14:06.1 & 0.209 & 0.343 & 0.147 & 0.337 & 0.031 &  sdM8.3 &  sdM7.9 &  sdL0.0  \cr
 9 & 08:43:58.50 & $+$06:00:38.6 & 0.814 & 0.551 & 0.377 & 0.544 & 0.281 &  sdM5.1 & esdM4.4 & usdM5.5  \cr
10 & 08:55:00.12 & $+$00:02:04.1 & 0.546 & 0.600 & 0.280 & 0.463 & 0.165 & esdM5.3 &  sdM5.7 &  sdM6.0  \cr
13 & 10:01:26.29 & $-$01:34:26.6 & 0.611 & 0.814 & 0.532 & 0.783 & 0.581 &   dM0.0 &   dM1.7 &  sdM5.0  \cr
16 & 11:58:26.62 & $+$04:47:46.8 & 0.140 & 0.816 & 0.218 & 0.387 & 0.088 &  sdM7.5 &   dM6.8 &  sdM9.5  \cr
17 & 12:02:14.62 & $+$07:31:13.8 & 0.959 & 0.314 & 0.258 & 0.439 & 0.137 & esdM5.6 & usdM6.1 & esdM7.0  \cr
17 & 12:02:14.62 & $+$07:31:13.8 & 0.877 & 0.720 & 0.401 & 0.511 & 0.275 & esdM4.6 & usdM4.4 & esdM7.0  \cr
18 & 12:15:08.37 & $+$04:02:00.5 & 0.528 & 0.628 & 0.218 & 0.405 & 0.097 & esdM6.0 &  sdM6.7 &  sdM7.0  \cr
20 & 12:36:59.43 & $-$00:21:58.2 & 0.693 & 0.609 & 0.343 & 0.558 & 0.268 & esdM4.0 & esdM4.5 & esdM5.5  \cr
21 & 12:42:34.62 & $+$14:33:06.2 & 0.985 & 0.556 & 0.342 & 0.521 & 0.242 & esdM4.5 & usdM4.8 & usdM5.0  \cr
22 & 12:44:25.90 & $+$10:24:41.9 & 0.135 & 0.657 & 0.180 & 0.272 & 0.018 &  sdM9.4 &   dM8.2 &  sdL0.5  \cr
23 & 12:46:21.90 & $+$04:43:09.9 & 0.702 & 0.688 & 0.421 & 0.625 & 0.372 &  sdM3.7 & esdM3.5 & esdM5.0  \cr
24 & 12:56:35.91 & $-$00:19:44.9 & 0.722 & 0.597 & 0.347 & 0.483 & 0.220 & esdM5.0 & esdM5.1 & esdM6.0  \cr
25 & 13:17:05.66 & $+$09:10:16.9 & 0.726 & 0.496 & 0.319 & 0.481 & 0.201 & esdM5.0 & esdM5.3 & esdM7.0  \cr
25 & 13:17:05.66 & $+$09:10:16.9 & 0.641 & 1.079 & 0.240 & 0.434 & 0.125 & esdM5.7 & esdM6.3 & esdM6.5  \cr
26 & 14:18:06.71 & $+$00:00:35.5 & 0.421 & 0.702 & 0.330 & 0.565 & 0.264 &  sdM4.6 &  sdM4.6 &  sdM6.5  \cr
27 & 14:54:41.42 & $+$12:35:56.7 & 0.742 & 0.348 & 0.499 & 0.633 & 0.231 & esdM4.8 & esdM4.9 &  esdM5.5 \cr
32 & 23:33:59.39 & $+$00:49:35.2 & 0.706 & 0.603 & 0.328 & 0.523 & 0.233 & esdM4.5 & esdM4.9 &  sdM6.0  \cr
  \hline
 \label{tab_subdw:tab_SpT_indices}
 \end{tabular}
\end{table*}
%

%
%%%%%%%%%%%%%%%%%%%%%%%%%%%%%%%%%%%%%%
%%%%% Figure: Spectral indices %%%%%
%%%%%%%%%%%%%%%%%%%%%%%%%%%%%%%%%%%%%%
%
%

%
%%%%%%%%%%%%%%%%%%%%%%%%%%%%%%%%%%%%%%%
%%%%% Figure: Spectral indices #2 %%%%%
%%%%%%%%%%%%%%%%%%%%%%%%%%%%%%%%%%%%%%%
%
% This plot was created with the IDL program by Marcela
%
%

%
%%%%%%%%%%%%%%%%%%%%%%%%%%%%%%%%%%%%%%
%%%%% Figure: SpType comparison %%%%%
%%%%%%%%%%%%%%%%%%%%%%%%%%%%%%%%%%%%%%
%
\begin{figure}
  \centering
  \includegraphics[width=\linewidth, angle=0]{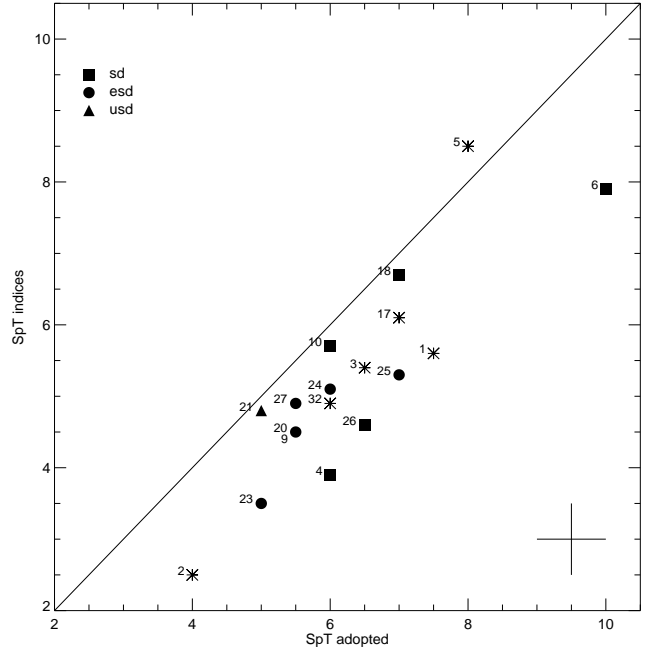}
  \caption{
Comparison of the spectral types obtained from the visual comparison
with metal-poor ``templates'' and spectral types derived from the
spectral indices defined by \citet{gizis97a} and \citet{lepine07c}.
Subdwarfs, extreme subdwarfs, and ultra-subdwarfs are marked as squares,
circles, and triangles, respectively. Metal-poor dwarfs with a discrepant
spectral types derived from both classification methods (comparison with
templates vs spectral indices) are displayed with
asterisks (e.g.\ sdM vs esdM or usdM vs esdM, etc).
Typical uncertainties on the spectral types derived from both methods
are half a subclass.
}
  \label{fig_subdw:fig_SpT_compare}
\end{figure}

The comparison between the spectral types inferred from spectral indices
and from the direct comparison with ``templates'' is shown in 
Fig.\ \ref{fig_subdw:fig_SpT_compare}. We find that the spectral types
derived from spectral indices tend to under-estimate the spectral type
(over-estimate the effective temperature). Therefore, we adopted the direct 
and visual comparison with ``templates'' to assign spectral types to our 
targets because it provides a more accurate classification.
We note that the spectral indices are not so reliable to classify subdwarfs
because they rely on a narrow wavelength range as pointed out by 
\citet{lepine07c} and also depend strongly on the resolution of the spectra
(Table \ref{tab_subdw:tab_SpT_indices}).
The differences between both classifications are listed in
Table \ref{tab_subdw:tab_SpT_indices} and shown in
Fig.\ \ref{fig_subdw:fig_SpT_compare} where asterisks represent confirmed
metal-poor dwarfs with discrepant classes (not only spectral types) derived 
from the spectral indices and the direct comparison with templates,
suggesting some shortcomings in the spectral types derived solely on indices.
Our sample of 20 low-mass stars with low-metal content consists of nine
subdwarfs, seven extreme subdwarfs, two ultra-subdwarfs, and two L subdwarfs.

The remaining seven sources in our spectroscopic sample are M dwarfs with
spectral types ranging from M4 to M7\@. We believe that the contamination 
of our sample by these M dwarfs come from large error bars on their
Sloan positions, leading to spurious proper motions placing them at the
bottom of the reduced proper motion diagram mimicking subdwarf candidates.
We remark that a revision of proper motions as that outlined in 
Section \ref{subdw:revised_PM} likely reduces the contamination level 
by a factor of 2--3\@. None of these M dwarfs
exhibit H$\alpha$ in emission at the resolution of our spectra (R$\sim$150),
suggesting that they are older than typical M dwarfs in the solar vicinity.
\citet{west08} reported H$\alpha$ in emission for 50\% of the sources or more 
at spectral types later than M4, with equivalent widths larger than 1\,\AA{}.
These authors also discussed the decrease in activity with age and scale height, 
suggesting that the M dwarfs contaminating our sample may be either older than 
the average low-mass stars in the solar neighbourhood and may be located at a
higher scale height. The lack of H$\alpha$ in emission also places a lower limit
on the ages of these M dwarfs, 5 and 8 Gyr for M5 and M7 dwarfs, respectively.

\subsection{Radial velocities}
\label{subdw:spec_RV}

In this section we tentatively compute the radial velocities of our targets
in spite of the low-resolution of our spectra. We used two different methods.

First, we computed the offsets between lines resolved in our FORS2 spectra and
the centroids of several atomic lines whose accurate positions can be found on
the webpage of the National Institute of Standards and Technology\footnote{
http://physics.nist.gov/asd3}: Ca\,{\small{I}} at 6572.18\,\AA{},  
Ti\,{\small{I}} at 8434.94\,\AA{}, Na\,{\small{I}} doublet at 8183.25 and 
8194.79\,\AA{}, Cs\,{\small{I}} at 8542.09 and 8943.47\,\AA{}, and 
Ca\,{\small{II}} at 8542.09\,\AA{} \citep[see also Table 2 of][]{burgasser09a}. 
The uncertainties on these offsets are typically of the order of 1\,\AA{}, 
corresponding to typical uncertainties of 35--50 km\,s$^{-1}$.
We focused only on the Ca\,{\small{I}}, Na\,{\small{I}} doublet, and
Ca\,{\small{II}} lines for the NOT spectrum and the SDSS spectra.
We derived the observed radial velocities by multiplying those offsets by the 
speed of light and dividing by the wavelength (c*$\Delta\lambda$/$\lambda$). 
Observed velocities were converted into heliocentric velocities by computing 
the Earth's rotation, the motion of the Earth's center about the Earth-Moon 
barycenter, and the motion of the Earth-Moon barycenter about the center of 
the Sun using the {\tt{RVCORRECT}} routine in IRAF. In the fifth column of 
Table \ref{tab_subdw:tab_RVs}  we provide our heliocentric radial velocities 
measured with this method.

Independently, we computed radial velocities by cross-correlating our optical
spectra with the optical spectrum of the sdL3.5 subdwarf 
SDSS J125637.13-022452.4 \citep{burgasser09a} using the task {\tt{FXCOR}} under
the IRAF environment. This subdwarf has a well-measured radial velocity
of $-$130 km/s and we decided to use it as a template despite having a
spectral type later than our confirmed subdwarfs. We transformed the vacuum
wavelength published by \citet{burgasser09a} into air wavelength and degraded
its resolution to match our observations. We cross-matched the full
spectrum of our sdM9.5 and sdL subdwarfs with Burgasser's template because of
the similarities in the shape of the spectra and spectral types but focused
only on the Na{\small{I}} doublet region for the earlier subdwarfs in our
sample. We cross-matched the spectra using two different functions: a gaussian 
fit to the cross-correlation function peak or simply the central value of the
peak. Both method led to differences smaller than
the uncertainties output by {\tt{FXCOR}}. The radial velocities inferred
by this method are listed in Table \ref{tab_subdw:tab_RVs} and have
typical uncertainties of 40--50 km\,s$^{-1}$ measured from the dispersion
of the measurements from different part of the spectra.
Radial velocities derived from both methods usually agree within the
uncertainties, except for ULAS J115826.62$+$044746.8 (ID=16) and
ULAS J12:36:59.43$-$00:21:58.2\@. In this case, 
we favour the radial velocities derived from the direct comparison with
SDSS J125637.13-022452.4\@.

%
%%%%%%%%%%%%%%%%%%%%%%%%%%%%%%%%%%%%%%%%%%
%%%%% Table Radial Velocities %%%%%
%%%%%%%%%%%%%%%%%%%%%%%%%%%%%%%%%%%%%%%%%%
%
\begin{table}
 \centering
 \caption[]{Coordinates (in J2000), spectral types, and radial velocities 
            (expressed in km\,s$^{-1}$) derived in two different ways: 
            first by measuring the offsets of well-known lines comparing 
            with their centers (typical uncertainties of 35--50 km\,s$^{-1}$)
            and, second, by cross-correlating the Na{\small{I}} doublet
            region of our optical spectra using SDSS J125637.13-022452.4 
            (sdL3.5) as template.}
 \begin{tabular}{@{\hspace{0mm}}c @{\hspace{2mm}}c @{\hspace{2mm}}c c c c@{\hspace{0mm}}}
 \hline
 \hline
ID &     R.A.    &    Dec        &   SpT   &  Vh$_{1}$  &      Vh$_{2}$  \cr
 \hline
   & hh:mm:ss.ss & $^{\circ}$:$'$:$''$ &     &  km\,s$^{-1}$ & km\,s$^{-1}$  \cr
 \hline
 1 & 00:45:39.97 & $+$13:50:32.7 &  sdM7.5 &     34.5  &      2.7  \cr
 3 & 01:33:46.25 & $+$13:28:22.4 & esdM6.5 &    107.1  &     29.0  \cr
 4 & 01:50:34.33 & $+$14:20:02.4 &  sdM6.0 &    205.7  &    134.2  \cr
 5 & 02:05:33.75 & $+$12:38:24.1 &  sdM8.0 &   $-$3.5  &      ---  \cr
 6 & 03:33:50.84 & $+$00:14:06.1 &  sdL0.0 &    251.4  &    368.0  \cr
 9 & 08:43:58.50 & $+$06:00:38.6 & usdM5.5 &    176.8  &    126.7  \cr
10 & 08:55:00.12 & $+$00:02:04.1 &  sdM6.0 &     69.7  &  $-$27.7  \cr
13 & 10:01:26.29 & $-$01:34:26.6 &  sdM5.0 &    115.8  &     82.8  \cr
16 & 11:58:26.62 & $+$04:47:46.8 &  sdM9.5 & $-$123.6  &      3.8  \cr
17 & 12:02:14.62 & $+$07:31:13.8 & esdM7.0 &    284.6  &    222.3  \cr
18 & 12:15:08.37 & $+$04:02:00.5 &  sdM7.0 &     51.8  &     76.7  \cr
20 & 12:36:59.43 & $-$00:21:58.2 & esdM5.5 &    219.5  &     68.9  \cr
21 & 12:42:34.62 & $+$14:33:06.2 & usdM5.0 &     97.6  &      7.4  \cr
22 & 12:44:25.90 & $+$10:24:41.9 &  sdL0.5 &    248.0  &    236.4  \cr
23 & 12:46:21.90 & $+$04:43:09.9 & esdM5.0 &    127.9  &    100.6  \cr
24 & 12:56:35.91 & $-$00:19:44.9 & esdM6.0 &    185.5  &    126.7  \cr
25 & 13:17:05.66 & $+$09:10:16.9 & esdM7.0 &     83.4  &     12.9  \cr
26 & 14:18:06.71 & $+$00:00:35.5 &  sdM6.5 & $-$115.8  & $-$161.0  \cr
27 & 14:54:41.42 & $+$12:35:56.7 & esdM5.5 &    128.6  &      --- \cr
32 & 23:33:59.39 & $+$00:49:35.2 &  sdM6.0 & $-$163.9  & $-$192.9  \cr
 \hline
 \label{tab_subdw:tab_RVs}
 \end{tabular}
\end{table}
%

%
%%%%%%%% Distance %%%%%%%%%%%
%
\subsection{Spectroscopic distances}
\label{subdw:spec_dist}

In this section, we estimate the spectroscopic distances by comparing our
new discoveries with subdwarfs of similar spectral types with known
trigonometric parallaxes.

We looked for subdwarfs with parallaxes whose spectral types are sdM5, sdM6,
sdM6.5, sdM7, sdM7.5, and sdM9.5\@. We considered the following sources
as subdwarfs with known distances to derive spectroscopic distances for our
new ultracool subdwarfs: 
LP\,807-23 \citep[sdM5.0; $J$ = 12.92 mag; d = 28.17--31.74 pc;][]{vanAltena95},
LHS\,1074 \citep[sdM6; $J$ = 14.68 mag; d = 85.7$\pm$17.1 pc;][]{salim03a,riaz08a},
LHS\,1166 \citep[sdM6.5; $J$ = 14.26 mag; d = 73--89 pc;][]{vanAltena95},
LP\,440-52 \citep[sdM7; $J$ = 13.19 mag; d = 34.4--36.1 pc;][]{vanAltena95},
LSR\,J203621.86$+$505950.3 \citep[sdM7.5; $J$ = 13.628 mag; d = 43.7--49.2 pc;][]{lepine02,schilbach09},
LSR\,J142504.81$+$710210.4 \citep[sdM8; $J$ = 14.828 mag; d = 75.4--89.9 pc;][]{lepine03d,burgasser08a,schilbach09}, and
SSSPM101307.34$-$135620.4 \citep[sdM9.5; $J$ = 14.637 mag; d = 45.0--54.6 pc;][]{scholz04c,schilbach09}.
After applying the standard transformations using the $J$-band magnitudes of
our new discoveries, we assigned mean distances between 88 and 628 pc (see
Table \ref{tab_subdw:tab_distances}), assuming that they are single. 
Typical error bars on the spectroscopic distances are 20--25\% taking into
account the uncertainties on the trigonometric parallaxes of the templates.
For the two L subdwarfs, we used the trigonometric parallax of the sdM9.5 
template to place upper limits on their distances. Both objects very likely 
lie within 100 pc unless they are binaries (Table \ref{tab_subdw:tab_distances}).

In addition to the aforementioned subdwarfs, we were able to assign a
spectroscopic distance for our esdM5 extreme subdwarf using LHS\,515 as template
\citep[esdM5; $J$ = 13.64 mag; d = 42.6--64.5 pc][]{vanAltena95}. Hence, we 
derive a spectroscopic distance of 257 pc with a probable range of 214--323 pc 
for ULAS J124621.90$+$044309.9 (ID=23; Table \ref{tab_subdw:tab_distances}).
For the other extreme subdwarfs we are unable to assign spectroscopic 
distances because no object with similar spectral types has known 
trigonometric parallax in the literature. Instead, we used 
LHS\,2096 \citep[esdM5.5; $J$ = 13.99; d = 56.10 pc;][]{lepine05d},
LHS\,2023 \citep[esdM6; $J$ = 14.91; d = 73.9 pc;][]{riaz08a},
LSR\,J0822$+$1700 \citep[esdM6.5; $J$ = 15.517; d = 106 pc;][]{lepine03c},
and APMPM0559 \citep[esdM7; $J$ = 14.887; d = 70 pc;][]{schweitzer99}
to infer tentative (mean) distances. We do not quote uncertainties for
these sources because of the (already) very uncertain distances of the
templates used. We do not provide a distance for the two ultra-subdwarfs in 
our sample because none has trigonometric parallaxes published in the literature.

%
%%%%%%%%%%%%%%%%%%%%%%%%%%%%%%%%%%%%%%%%%%
%%%%% Table Distances %%%%%
%%%%%%%%%%%%%%%%%%%%%%%%%%%%%%%%%%%%%%%%%%
%
\begin{table}
 \centering
 \caption[]{Coordinates (in J2000), spectral types, and spectroscopic distance
            estimates (in pc) for our new subdwarfs. 
}
 \begin{tabular}{c c c c c}
 \hline
 \hline
ID &     R.A.    &  Dec        & SpT & Distance  \cr
 \hline
   & hh:mm:ss.ss & $^{\circ}$:$'$:$''$ &     &  pc \cr
 \hline
 1 & 00:45:39.97 & $+$13:50:32.7 &  sdM7.5 & 354 (334--376) \cr
 3 & 01:33:46.25 & $+$13:28:22.4 & esdM6.5 & 124$^{a}$ \cr
 4 & 01:50:34.33 & $+$14:20:02.4 &  sdM6.0 & 303 (243--364) \cr
 5 & 02:05:33.75 & $+$12:38:24.1 &  sdM8.0 & 132 (121-146)  \cr
 6 & 03:33:50.84 & $+$00:14:06.1 &  sdL0.0 & $<$97 (88--107)  \cr
 9 & 08:43:58.50 & $+$06:00:38.6 & usdM5.5 & no template    \cr
10 & 08:55:00.12 & $+$00:02:04.1 &  sdM6.0 & 524 (419--628) \cr
13 & 10:01:26.29 & $-$01:34:26.6 &  sdM5.0 & 464 (438--494) \cr
16 & 11:58:26.62 & $+$04:47:46.8 &  sdM9.5 & 116 (105--128) \cr
17 & 12:02:14.62 & $+$07:31:13.8 & esdM7.0 & 237$^{a}$ \cr
18 & 12:15:08.37 & $+$04:02:00.5 &  sdM7.0 & 281 (274--288) \cr 
20 & 12:36:59.43 & $-$00:21:58.2 & esdM5.5 & 171$^{a}$ \cr
21 & 12:42:34.62 & $+$14:33:06.2 & usdM5.0 & no template \cr
22 & 12:44:25.90 & $+$10:24:41.9 &  sdL0.5 & $<$104 (95--115)  \cr
23 & 12:46:21.90 & $+$04:43:09.9 & esdM5.0 & 257 (214--323) \cr
24 & 12:56:35.91 & $-$00:19:44.9 & esdM6.0 & 132$^{a}$ \cr
25 & 13:17:05.66 & $+$09:10:16.9 & esdM7.0 & 214$^{a}$ \cr
25 & 13:17:05.66 & $+$09:10:16.9 & esdM6.5 & 242$^{a}$ \cr
26 & 14:18:06.71 & $+$00:00:35.5 &  sdM6.5 & 464 (424--517) \cr
27 & 14:54:41.42 & $+$12:35:56.7 & esdM5.5 & 150$^{a}$      \cr
32 & 23:33:59.39 & $+$00:49:35.2 &  sdM6.0 & 399 (319--478) \cr
 \hline
 \label{tab_subdw:tab_distances}
 \end{tabular}
\vspace{-0.25cm}
\newline
$^{a}$ the range in distances is not listed because the templates used do 
not have  trigonometric parallaxes.
\end{table}
%

%
%%%%%%%%%%%%%%%%%%%%%%%%%%%%%%%%%%%%%%%%%%
%%%%% Table WISE photometry %%%%%
%%%%%%%%%%%%%%%%%%%%%%%%%%%%%%%%%%%%%%%%%%
%
\begin{table*}
 \centering
 \caption[]{Coordinates (in J2000), spectral type, $J$-band magnitude, and
           mid-infrared photometry with the associated error bars (3.4 and 4.6 
           microns) for the two ultracool subdwarfs covered by the WISE mission. 
           The signal-to-noise ratio of the photometry quoted by the WISE 
           catalogue is indicated in brackets. The photometry in the WISE 12
           and 22 micron bands is not included because the signal-to-noise
           ratios are less than 3\@.}
 \begin{tabular}{c c c c c c c}
 \hline
 \hline
ID & R.A.  & Dec  &  SpType & $J$ & 3.4$\mu$m (SNR) & 4.6$\mu$m (SNR) \\
 \hline
   & hh:mm:ss.ss  & $^{\circ}$:$'$:$''$ &   & mag & mag$\pm$err () & mag$\pm$err () \\  
 \hline
 4 & 01:50:34.33 & $+$14:20:02.4 &  sdM6.0 & 17.424 & 16.667$\pm$0.131 ( 8.3) & 15.877$\pm$0.230 ( 4.7)  \\ 
25 & 14:54:41.42 & $+$12:35:56.7 &  sdM5.5 & 16.121 & 15.136$\pm$0.044 (24.4) & 14.837$\pm$0.090 (12.1)  \\ 
  \hline
 \label{tab_subdw:tab_WISE_phot}
 \end{tabular}
\end{table*}
%

%
%%%%%%%% WISE mid-IR photometry %%%%%%%%%%%
%
\subsection{Mid-infrared photometry}
\label{subdw:spec_midIR}

\begin{figure}
  \centering
  \includegraphics[width=\linewidth, angle=0]{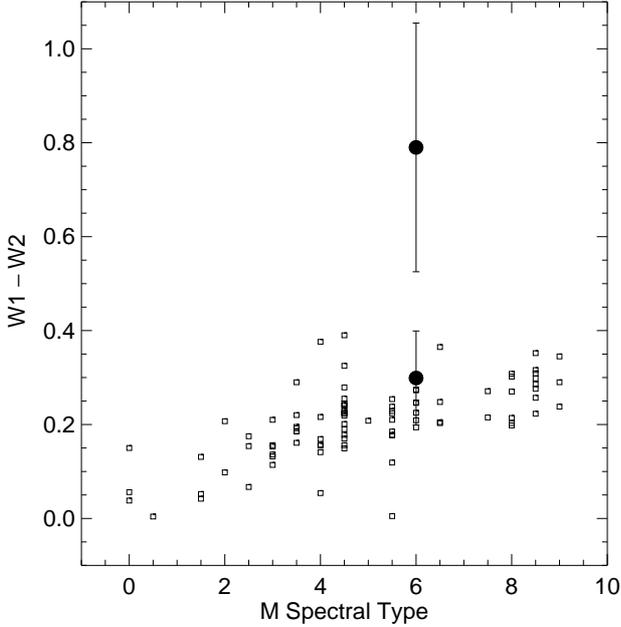}
  \caption{
WISE colours as a function of spectral type for known M dwarfs (open squares) 
from the sample published by \citet{kirkpatrick11}. Our two subdwarf candidates 
are highlighted with large filled circles.
}
  \label{fig_subdw:fig_SpT_WISEcol}
\end{figure}

We cross-matched our list of new ultracool subdwarfs with the Wide-field Infrared
Survey Explorer \citep[WISE;][]{wright10} data release which took place in April 
2011\@. We found two subdwarfs with mid-infrared photometry using a matching 
radius of 6.5 arcsec, the spatial resolution of WISE\@. However, we ensured that 
no other source was detected in WISE because of the large proper motions of 
subdwarfs. The WISE photometry of these two subdwarfs at 3.4 (W1) and 4.6 (W2)
microns is reported in Table \ref{tab_subdw:tab_WISE_phot} and plotted in 
Fig.\ \ref{fig_subdw:fig_SpT_WISEcol}; these two sources are undetected
at longer wavelengths.

We compared the infrared colours of our new ultracool subdwarfs to the sample
of \citet{kirkpatrick11} drawn from the DwarfArchives.org 
webpage\footnote{http://spider.ipac.caltech.edu/staff/davy/ARCHIVE/index.shtml}
in order to identify any trend that may help future searches for metal-poor 
brown dwarfs. 
In the case of ULAS J015034.33$+$142002.4 (ID=4; sdM6), we find colours 
similar to normal M6 dwarfs in $H-$W2 = 1.07, $J-$W2=0.76, and $K-$W1 although 
on the red side of the distribution. Other colours, including $J-$W1 = 0.76, 
W1$-$W2 = 0.79, $K-$W2 = 0.92 mag differ to those of normal M6 dwarfs which 
span the following ranges 1.0--1.4, 0.2--0.4, and 0.3--0.6 mag, resulting in a
deviation of $\sim$1$\sigma$ (Fig. \ref{fig_subdw:fig_SpT_WISEcol}; 
Table \ref{tab_subdw:tab_WISE_phot}). The other subdwarf in our sample, 
ULAS145441.42$+$123556.7 (ID=27; sdM5.5), exhibits 
similar colours to M dwarfs in all combinations of colours at odds with
ULAS J015034.33$+$142002.4 (Fig.\ \ref{fig_subdw:fig_SpT_WISEcol}). We repeated 
the process with a larger number of known subdwarfs and did not spot any obvious 
trend with decreasing metallicity, suggesting that ULAS J015034.33$+$142002.4 
may be peculiar, or a multiple source, or more likely the WISE photometry has 
underestimated uncertainties. 

We also checked the Spitzer archive to extract additional information on the 
mid-infrared properties of some of our candidates but none of them was found
from the Spitzer public database.

%
%%%%%%%%%%%%%%%%%%%%%%%%%%%%%%%%%%%%%%
%%%%% Figure: WISE photometry %%%%%
%%%%%%%%%%%%%%%%%%%%%%%%%%%%%%%%%%%%%%
%
%

%
%%%%%%%% Wide companions %%%%%%%%%%%
%
\subsection{Search for wide companions}
\label{subdw:spec_WideComp}

In this section we looked for wide companions brighter than each of our subdwarfs
within a radius of 10 arcmin. The idea is to find potential primaries with
distances, metallicities, and (possibly) ages \citep[often refered to as 
benchmark objects;][]{pinfield06} determined with a higher precision than our 
new subdwarfs. We used large-scale surveys with accurate proper motions: 
the USNO-B1 \citep{monet03}, the UCAC3 \citep{zacharias10a}, and the PPMXL 
\citep{roeser10} catalogues. We selected only bright sources with 
$I_{\rm USNO}$\,$\leq$\,18 mag and proper motions in right ascension and 
declination within 30\% of the measured motion of our targets derived from 
the UKIDSS LAS DR5 vs SDSS DR7 cross-match.

We found one potential wide companion to ULAS J233359.39$+$004935.2 (ID=32) 
in USNO-B1 and PPMXL at a distance of about 5 arcmin on the sky. This potential 
companion, USNO J233350.27$+$005342.9, has a proper motion in right ascension 
and declination of (149, $-$75) and (154, $-$78) mas/yr in PPXML and USNO-B1, 
respectively, in agreement with the values of (136, $-$71) mas/yr quoted by 
Sloan. The differences in the proper motions in right ascension and in
declination between our subdwarf and its potential companion are about 
23--27\% and 10--15\%, respectively. No spectrum is available in the SDSS 
spectroscopic database. The SDSS$i$ of this object is 19.118$\pm$0.021 mag, 
roughly 1 mag brighter than our spectroscopic subdwarf. Its optical colours 
($g-r$ = 1.785, $r-i$ = 1.429, $r-z$ = 2.158 mag) and reduced proper motion
(H$_{r}$ = 21.64 mag) satisfy our original selection criteria and suggests that 
this potential wide companion may also be metal-poor. However, its $J-K$ 
infrared are redder than our original cut of 0.7 mag, implying that this object 
felt out of our sample. At the spectroscopic distance of 
ULAS J233359.39$+$004935.2 (ID=32) estimated to 400\,pc 
(Table \ref{tab_subdw:tab_distances}), the projected physical separation of 
the system would be very larger, of the order of 120,000 au. Hence, we cannot 
claim that both objects are gravitationally bound but they might have formed 
in the same cluster or might belong to the same moving group.

%
%%%%%%%% Notes on individual objects %%%%%%%%%%%
%
\subsection{Notes on individual objects}
\label{subdw:spec_notes}

In this section we give additional details on a few specific candidates
identified in our search for ultracool subdwarfs.

\subsubsection{Spectra in the SDSS DR7 spectroscopic database}

Some photometric candidates in our sample have spectra publicly available in
the SDSS DR7 spectroscopic database (Table \ref{tab_subdw:tab_log_spectro}). 
As noted in the caption of Table \ref{tab_subdw:tab_SDSS_UKIDSS} in Appendix,
all of them were included in the sample of \citet{west08} and classified
as early-M dwarfs using the Hammer classification \citep{covey07}.
However, the Sloan spectra clearly look like subdwarf with strong CaH
and TiO bands typical of low-metallicity M dwarfs 
(Figs.\ \ref{fig_subdw:fig_sdM}--\ref{fig_subdw:fig_esdM}).
We should mention that LP\,468-277 (01:33:46.25$+$13:28:22.4) was included
in the catalogue of Northern stars of \citet{lepine05d} but no spectral type 
was derived.
Finally, we recovered SDSS J020533.75$+$123824.1 (ID=5) classified as sdM7.5 by
\citet{lepine08b} and re-classified as sdM8 in this paper based on the 
spectral type provided by the SDSS DR7 spectroscopic database.

\subsubsection{ULAS J145441.45$+$123557.6 (ID\,=\,27)}

This source is part of the catalogue of Northern stars of \citet{lepine05d}
but no spectral type was derived. Our proper motion derived from the LAS
and SDSS DR7 epochs (0.31 arcsec/yr) is in good agreement with the value of
0.321 arcsec/yr published by \citet{lepine05d}. We do not have spectroscopic
follow-up for this source yet.

\subsubsection{ULAS J233359.39$+$004935.2 (ID\,=\,32)}

This candidate, observed with VLT FORS2, is classified as a sdM6 subdwarf.
Its proper motion in right ascension and declination is (93.9, 23.0) and
(81.9, 16.6) mas/yr from the 2MASS/UKIDSS and SDSS/UKIDSS cross-match.
We discuss the presence of a possible wide companion five arcmin away in
Section \ref{subdw:spec_WideComp}.

\begin{figure}
  \centering
  \includegraphics[width=\linewidth, angle=0]{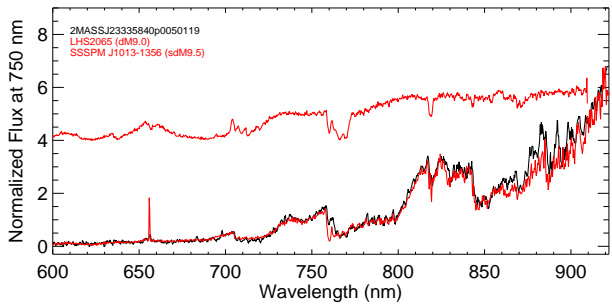}
  \caption{
Comparison of 2MASS J233358.40$+$005011.9 classified as L0 by \citet{zhang10} 
with a known solar-metallicity, LHS\,2065, classified as M9 by 
\citet{kirkpatrick91} and a sdM9.5 subdwarf \citep{scholz04c}.
}
  \label{fig_subdw:fig_L0_comparison}
\end{figure}

This subdwarf is located at $\sim$42 arcsec from 2MASS J233358.40$+$005011.9,
a L0 dwarf reported by \citet{zhang10} with a proper motion of (139.7, 29.5)
mas/yr reported by the PPMXL catalogue \citep{roeser10}.
The proper motion values in right ascension differ by 50--70\%, suggesting
that both objects are not physically associated. Nonetheless, we investigated 
further the SDSS DR7 spectrum of this possible wide companion and compared
it to known M9 and M9.5 dwarfs, young L dwarf templates \citep{cruz09}, a
sdM9.5 subdwarf \citep{scholz04c}, and our two possible L subdwarfs
(Fig.\ \ref{fig_subdw:fig_L0_comparison}). The best fit is obtained for a field 
M9 dwarf of solar metallicity, LHS\,2065 \citep{kirkpatrick91} downloaded from 
Kelle Cruz's webpage (Fig.\ \ref{fig_subdw:fig_L0_comparison})\footnote{http://kellecruz.com/M\_standards/}, earlier by
one subclass compared to the classification by \citet{zhang10}.

Overall conclusion: these two sources separated by about 42 arcsec are
unlikely to be physically associated. Note that we also discussed in the 
previous section (Sect.\ \ref{subdw:spec_WideComp}) that this object may 
be part of a wide binary system or a old moving group.

%
%%%%%%%%%%%%%%%%%%%%%%%%%%%%%%%%%%%%%%%
%%%%% Figure: L0 dwarf comparison %%%%%
%%%%%%%%%%%%%%%%%%%%%%%%%%%%%%%%5%%%%%%
%
\subsubsection{ULAS J033350.84$+$001406.1 (ID\,=\,6)}

This bright source ($J$ = 16.1 mag) has a large proper motion of 0.78 arcsec/yr. 
Its VLT FORS2 spectrum appears slightly redder than ULAS J115826.62$+$044746.8
(ID=16) which is well reproduced by the IRTF/SpeX spectrum of SSSPM\,1013$-$1356 
\citep[sdM9.5;][]{scholz04c,burgasser04}. Hence, we propose this object as a 
new sdL0 template for future searches because it is the first of its class
known to date.

\subsubsection{ULAS J124425.90$+$102441.9 (ID\,=\,22)}

This bright ($J$ = 16.26 mag) candidate is the coolest object in our sample, 
cooler than ULAS J033350.84$+$001406.1 (ID=6; see previous section).
Hence we propose it as a sdL0.5 template for future works on subdwarfs
because it is the first of its class at the time of writing.
Its proper motion also exceeds the 0.5 arcsec/yr threshold of high proper
motion sources, with a total motion of 0.67 arcsec/yr.
We note that the two new targets classified as L subdwarfs occupy the
lower right of the reduced proper motion diagram (Fig.\ \ref{fig_subdw:fig_RPM}),
suggesting a steep reddening in the colours of subdwarfs in the M/L transition.

%
%%%%%%%% Surface density %%%%%%%%%%%
%
\subsection{Surface density of subdwarfs}
\label{subdw:spec_density}

In this section, we provide a tentative estimate of the surface density of 
ultracool subdwarfs i.e.\ metal-poor dwarfs with spectral types later than 
M5\@. We found a total of 18 M5--M9.5 subdwarfs (and two additional subdwarfs
classified as L0--L0.5) over 1343 square degrees common to UKIDSS LAS DR5
and SDSS DR7 and imaged in the $YJHK$ filters. We should potentially add 
one source without spectroscopy yet. We derive a density of 
0.015--0.016 $\geq$M5 subdwarfs per square degree.

\citet{west11} identified 70,841 spectroscopic M0--M9 dwarfs in the 8200 square
degrees covered by the SDSS spectroscopic database. 
However the SDSS spectroscopic follow-up of the imaging survey
is incomplete. For this reason, we considered the photometric sample from 
\citet[][their Figure 4]{bochanski10} and focus on sources with $i$ brighter
than 22 mag, $r-z$\,$\geq$\,2.5, $i-z$\,$\geq$\,0.2, and $r-i$\,$\geq$\,0.3 mag,
corresponding to $\geq$\,M5 dwarfs and later. We counted a total number of 
653,625 photometric $\geq$\,M5 dwarfs in 8000 square degrees surveyed by
SDSS DR5\footnote{Numbers kindly provided by John Bochanski and Andrew West},
implying a density of $\sim$82 $\geq$\,M5 dwarfs per square degree. We sent a 
query with the aforementioned criteria to the WFCAM Science Archive 
\citep{hambly08} to see how many M5 (and later) dwarfs in UKIDSS LAS DR5 and 
SDSS DR7 we could recover in order to match those numbers with our search 
criteria defined in Sect.\ \ref{subdw:select}. We imposed a detection in $YJH$ 
but not in $K$ and requested good quality point sources in addition to the
constraints on the optical colours. The query returned 113,393 sources in 1343
square degrees. If we limit the sample to dwarfs with $z-J$ greater than 
1.4 \citep{west11}, the query returns 106,746 sources, implying a number of
M5 dwarfs (and later) of the order of 79.5--84.4 per square degree which is 
highly consistent with the numbers derived from the Sloan sample alone.
This density is $\sim$5000--5700 times higher than the number of ultracool 
subdwarfs found in our photometric and proper motion search which is broadly
consistent with the 0.2\% contribution from metal-poor stars quoted by
\citet{digby03} and the upper limit derived from the SDSS M dwarf sample 
\citep{covey08a}. We should mention that according to the evolutionary
models \citep{baraffe97,baraffe98}, the masses of $\geq$\,M5 dwarfs and 
subdwarfs are similar at ages of Gyr but lower metallicity M dwarfs have high 
effective temperatures. Finally, we should point out that we found two 
ultra-subdwarf and seven extreme subdwarfs for 11 subdwarfs, suggesting a 
fairly quick decrease in the numbers of subdwarfs as a function of metallicity.

%
%%%%%%%%%%%%%%%%%%%%%%%%%%%%%%%%%%%%%
%%%%%%%  SUMMARY %%%%%%%
%%%%%%%%%%%%%%%%%%%%%%%%%%%%%%%%%%%%%
%
\section{Summary}
\label{subdw:summary}

We have presented the outcome of a dedicated photometric and proper motion 
search aimed at finding new ultracool subdwarfs in public databases. We 
identified 32 ultracool subdwarf candidates, 20 of them being confirmed as 
metal-poor late-M and early-L dwarfs by low-resolution optical spectroscopy. 
We discovered two new early-L subdwarfs which we propose as spectroscopic 
templates for future searches because these are the first of their subclass. 
We measured radial velocities for most of the new subdwarfs with the 
cross-match technique. We estimated their spectroscopic distances when 
templates of similar spectral types with trigonometric parallaxes were 
available. We uncovered seven 
old M dwarfs contaminating our sample whose ages are estimated to $>$5--8 Gyr 
due to the lack of H$\alpha$ in emission. Of the 32 candidates, five 
do not have optical spectroscopy. Only one of these five remains a good 
subdwarf candidate, the others being rejected. We found a contamination 
of about 30\% by solar-metallicity M dwarfs in our photometric and proper 
motion search, mainly due to large errors on the Sloan positions 
leading to spurious proper motions affecting the determination of the 
reduced proper motion. We are able to reduce this level of contamination
by a factor 2 to 3 after revision of the proper motion measurements.
We also present mid-infrared data from WISE for two subdwarfs as well as a 
search for bright and wide common proper motions which led to an extremely 
wide pair very likely not gravitationally bound.

We intend to expand our search for subdwarfs with upcoming data releases from
UKIDSS to increase the census of low-metallicity dwarfs. Moreover, we plan to
update our colour criteria to optimize future searches and discover even
cooler ultracool subdwarfs. The main overall scientific goal of this large 
project is to update and extend the current low-metallicity
classification into the L dwarf (and later T dwarf) regime.

%
%%%%%%%%%%%%%%%%%%%%%%%%%%
%%%%% Acknowledgments %%%%%
%%%%%%%%%%%%%%%%%%%%%%%%%%
%
\begin{acknowledgements}
NL was funded by the Ram\'on y Cajal fellowship number 08-303-01-02 and the 
national program AYA2010-19136 funded by the Spanish ministry of science and 
innovation. NL was partly funded by the RoPACS (Rocky Planets Around Cool Stars)
Marie Curie Initial Training Network. This research has made use of the Spanish 
Virtual Observatory supported from the Spanish MICINN through grant AYA2008-02156\@. 
Additional support was provided by the CONSOLIDER-INGENIO GTC project. 
We thank John Bochanski and Andrew West who kindly provided the numbers
of M dwarfs found in Sloan. We would like to thank the anonymous referee
for her/his valuable comment on proper motions.

The data presented here were obtained [in part] with ALFOSC, which is provided 
by the Instituto de Astrof\'isica de Andalucia (IAA) under a joint agreement 
with the University of Copenhagen and the NBIfAFG of the Astronomical Observatory of Copenhagen.
This article is partly based on observations obtained in service mode with the 
Nordic Optical Telescope, operated on the island of La Palma jointly by Denmark, 
Finland, Iceland, Norway, and Sweden, in the Spanish Observatorio del Roque de 
los Muchachos of the Instituto de Astrof\'isica de Canarias. We would like to 
thank Jorge Garcia Rojas and the entire Support Astronomer Group for taking 
data in service mode for us.

This work made use of data taken within the framework of the UKIRT Infrared 
Deep Sky Survey (UKIDSS). The United Kingdom Infrared Telescope is operated 
by the Joint Astronomy Centre on behalf of the U.K.\ Science Technology and
Facility Council. 

This research has made use of the Simbad and Vizier databases, operated
at the Centre de Donn\'ees Astronomiques de Strasbourg (CDS), and
of NASA's Astrophysics Data System Bibliographic Services (ADS).

This publication makes use of data products from the Two Micron
All Sky Survey (2MASS), which is a joint project of the University
of Massachusetts and the Infrared Processing and Analysis
Center/California Institute of Technology, funded by the National
Aeronautics and Space Administration and the National Science Foundation.

Funding for the SDSS and SDSS-II has been provided by the Alfred P. Sloan 
Foundation, the Participating Institutions, the National Science Foundation, 
the U.S. Department of Energy, the National Aeronautics and Space Administration, 
the Japanese Monbukagakusho, the Max Planck Society, and the Higher Education 
Funding Council for England. The SDSS Web Site is http://www.sdss.org/.
The SDSS is managed by the Astrophysical Research Consortium for the 
Participating Institutions. The Participating Institutions are the American 
Museum of Natural History, Astrophysical Institute Potsdam, University of Basel, 
University of Cambridge, Case Western Reserve University, University of Chicago, 
Drexel University, Fermilab, the Institute for Advanced Study, the Japan 
Participation Group, Johns Hopkins University, the Joint Institute for Nuclear 
Astrophysics, the Kavli Institute for Particle Astrophysics and Cosmology, the 
Korean Scientist Group, the Chinese Academy of Sciences (LAMOST), Los Alamos 
National Laboratory, the Max-Planck-Institute for Astronomy (MPIA), the 
Max-Planck-Institute for Astrophysics (MPA), New Mexico State University, Ohio 
State University, University of Pittsburgh, University of Portsmouth, Princeton 
University, the United States Naval Observatory, and the University of Washington.

The DENIS project has been partly funded by the SCIENCE and the HCM plans
of the European Commission under grants CT920791 and CT940627.
It is supported by INSU, MEN and CNRS in France, by the State of
Baden-W\"urttemberg in Germany, by DGICYT in Spain, by CNR in Italy,
by FFwFBWF in Austria, by FAPESP in Brazil, by OTKA grants F-4239 and
F-013990 in Hungary, and by the ESO C\&EE grant A-04-046.
Jean Claude Renault from IAP was the Project manager. Observations were
carried out thanks to the contribution of numerous students and young
scientists from all involved institutes, under the supervision of
P.\ Fouqu\'e, survey astronomer resident in Chile.

This research has made use of data obtained from the SuperCOSMOS
Science Archive, prepared and hosted by the Wide Field Astronomy Unit,
Institute for Astronomy, University of Edinburgh, which is funded by
the UK Particle Physics and Astronomy Research Council.

This publication makes use of data products from the Wide-field Infrared
Survey Explorer, which is a joint project of the University of California,
Los Angeles, and the Jet Propulsion Laboratory/California Institute of
Technology, funded by the National Aeronautics and Space Administration.

This research has benefitted from the M, L, and T dwarf compendium housed at 
DwarfArchives.org and maintained by Chris Gelino, Davy Kirkpatrick, and Adam 
Burgasser. This research has benefitted from the SpeX Prism Spectral Libraries, 
maintained by Adam Burgasser at http://www.browndwarfs.org/spexprism.
This research has also benefitted from the M dwarf standard spectra made
available by Kelle Cruz at www.astro.caltech.edu/\~kelle/M\_standards/.

\end{acknowledgements}

%
%%%%%%%%%%%%%%%%%%%%%%%%%%
%%%%% Bibliography %%%%%
%%%%%%%%%%%%%%%%%%%%%%%%%%
%
% for the bibliography
  \bibliographystyle{aa}
  \bibliography{../mnemonic,../biblio_old}

\appendix
%
%%%%%%%%%%%%%%%%%%%%%%%%%%%%%%%%%%%%%%%
%%%%% Figure: Finding chart #1 %%%%%
%%%%%%%%%%%%%%%%%%%%%%%%%%%%%%%%%%%%%%%
%
\begin{figure*}
  \centering
  \includegraphics[width=\linewidth, angle=0]{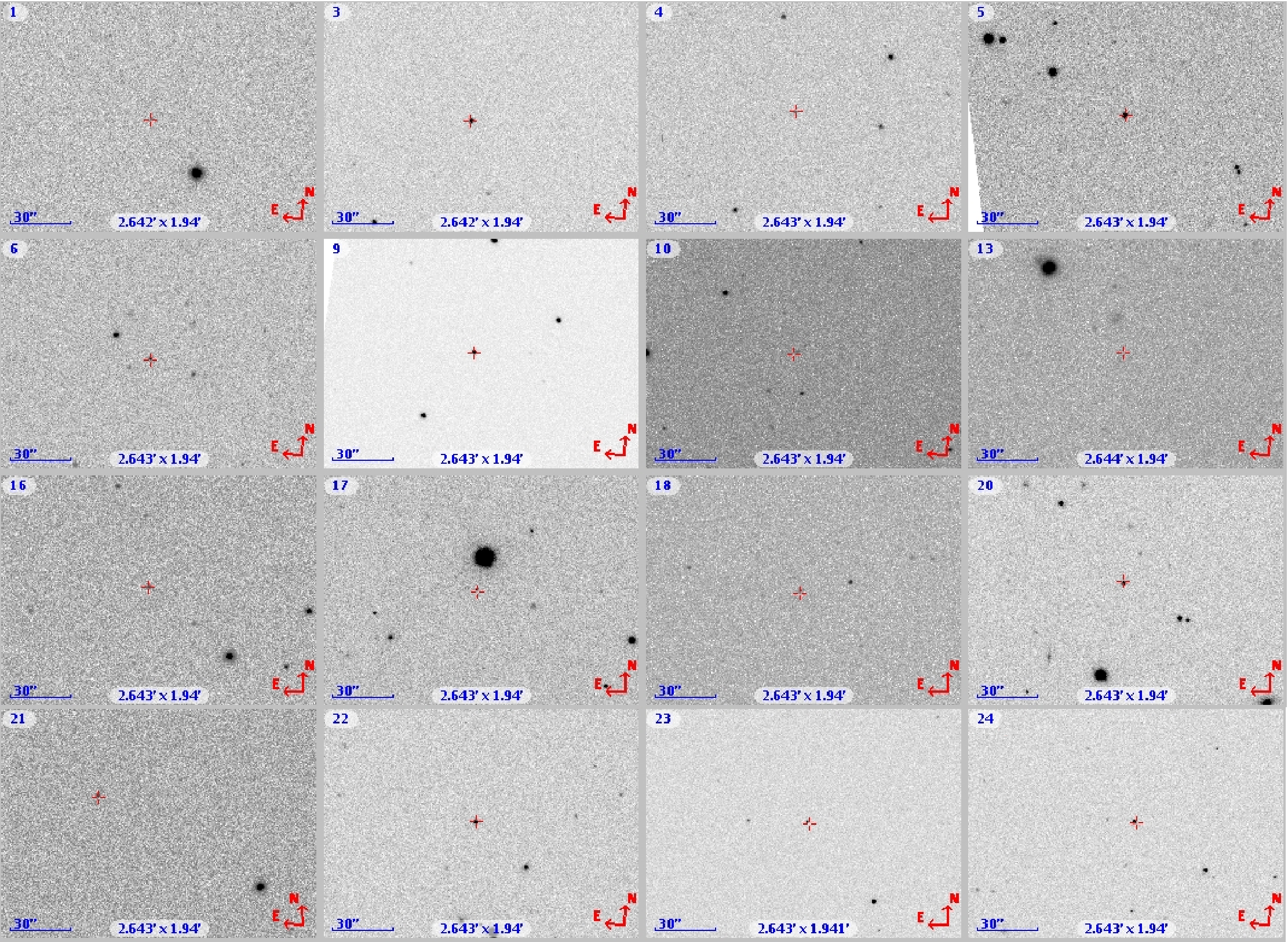}
  \caption{
Finding charts for 16 confirmed subdwarfs. Images are in the Sloan
$z$-band and approximately 2.64$\times$1.94 arcmin aside with North up 
and East left. The ID number of the object is indicated at the top left
of the image and the scale at the bottom left.
}
  \label{fig_subdw:fc_subdw_1}
\end{figure*}
%

%
%%%%%%%%%%%%%%%%%%%%%%%%%%%%%%%%%%%%%%%
%%%%% Figure: Finding chart #2 %%%%%
%%%%%%%%%%%%%%%%%%%%%%%%%%%%%%%%%%%%%%%
%
\begin{figure*}
  \centering
  \includegraphics[width=\linewidth, angle=0]{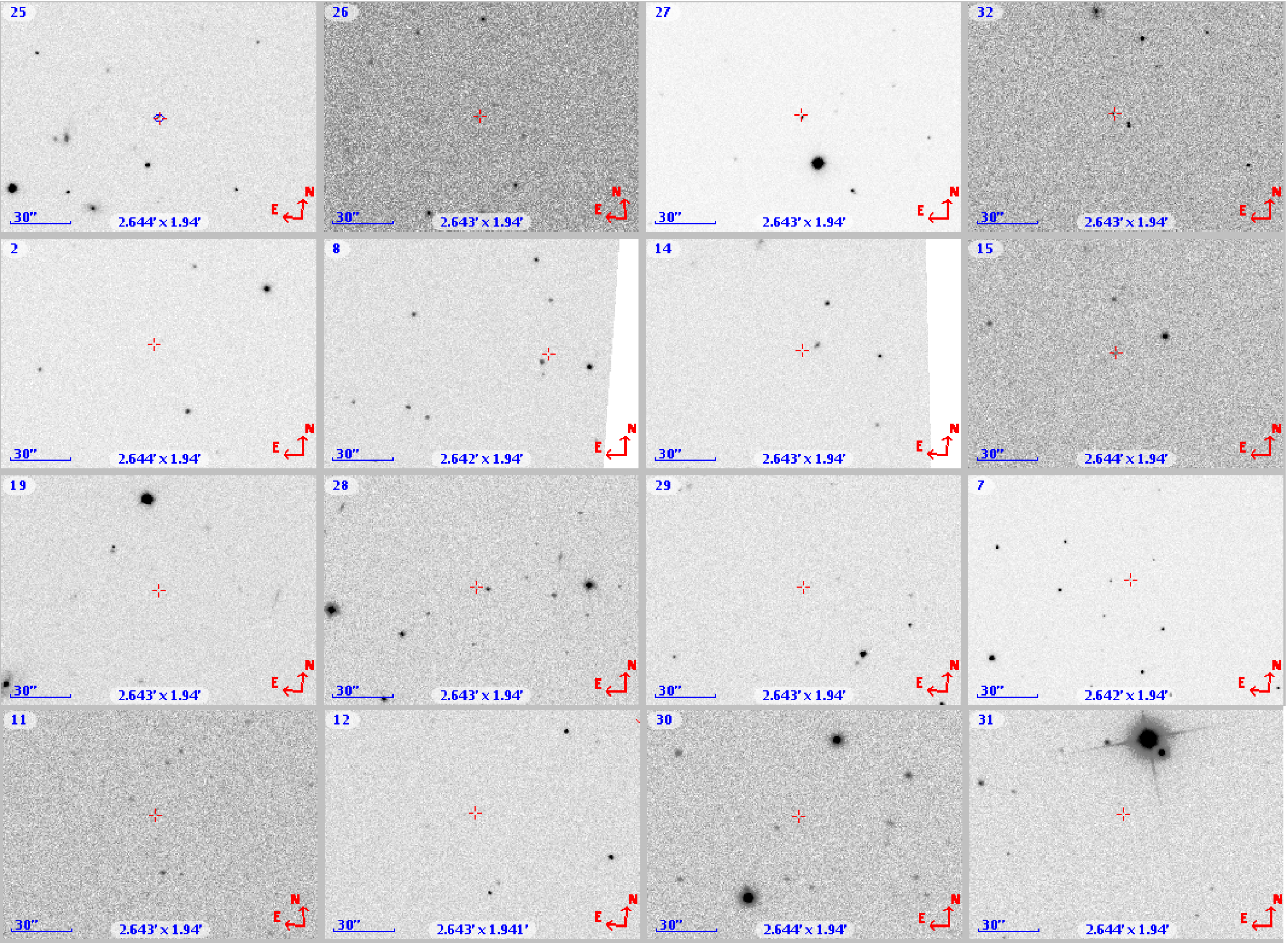}
  \caption{
The other 16 candidates identified in this work: the remaining four subdwarfs 
confirmed spectroscopically are shown at the top, the seven solar-metallicity
M dwarfs below followed by the five candidates without optical spectra.
Images are in the Sloan
$z$-band and approximately 2.64$\times$1.94 arcmin aside with North up 
and East left. The ID number of the object is indicated at the top left
of the image and the scale at the bottom left.
}
  \label{fig_subdw:fc_subdw_2}
\end{figure*}
\end{document}